\documentclass{ptephy}

\usepackage{amsmath} 
\usepackage{graphics} 

\usepackage{graphicx}
\usepackage{amssymb}
\makeatletter

 \@addtoreset{equation}{section}
\makeatother
\newcommand{\beq}{\begin{eqnarray}}
\newcommand{\eeq}{\end{eqnarray}}
\usepackage{color}

\begin{document}

\title{Properties of the twisted Polyakov loop coupling and the infrared fixed point in the SU(3) gauge theories}

\author{\name{Etsuko Itou}{1}}

\address{\affil{1}{High Energy Accelerator Research Organization (KEK), Tsukuba 305-0801, Japan}
\email{eitou@post.kek.jp}}

\begin{abstract}%
We report the nonperturbative behavior of the twisted Polyakov loop (TPL) coupling constant for the SU($3$) gauge theories defined by the ratio of Polyakov loop correlators in finite volume with twisted boundary condition.
We reveal the vacuum structures and the phase structure for the lattice gauge theory with the twisted boundary condition.
Carrying out the numerical simulations, we determine the nonperturbative running coupling constant in this renormalization scheme for the quenched QCD and $N_f=12$ SU($3$) gauge theories.

At first, we study the quenched QCD theory using the plaquette gauge action.
The TPL coupling constant has a fake fixed point in the confinement phase.
We discuss this fake fixed point of the TPL scheme and obtain the nonperturbative running coupling constant in the deconfinement phase, where the magnitude of the Polyakov loop shows the nonzero values.

We also investigate the system coupled to fundamental fermions.
Since we use the naive staggered fermion with the twisted boundary condition in our simulation, only multiples of $12$ are allowed for the number of flavors.
According to the perturbative two loop analysis, the $N_f=12$ SU($3$) gauge theory might have a conformal fixed point in the infrared region. 
However, the recent lattice studies show controversial results for the existence of the fixed point.
We point out possible problems in previous works, and present our careful study.
Finally, we find the infrared fixed point (IRFP) and discuss the robustness of the nontrivial IRFP of many flavor system under the change of the analysis method.

A part of preliminary results was reported in the proceedings \cite{Bilgici:2009nm,Itou:2010we} and the letter paper \cite{Aoyama:2011ry}.
In this paper we include a review of these results and give a final conclusion for the existence of IRFP of SU($3$) $N_f=12$ massless theory using the updated data.
\end{abstract}

\subjectindex{B32, B38, B44}

\maketitle

\section{Introduction}
Lattice gauge theory is one of the most successful regularization tools to understand QCD.
Since it can be applied even to the strong coupling region, we can investigate the nonperturbative properties of QCD.
Based on the success of the lattice QCD, there are several application to the lattice gauge theories with different gauge groups, numbers of flavors and fermion representations.
Among recent lattice studies, the search for the conformal or nearly conformal field theory in the infrared (IR) regime has been performed, motivated by both theoretical and phenomenological interests~\cite{Iwasaki:2003de} -- \cite{Karavirta:2012qd}.
If there is an IRFP with nonzero gauge coupling constant, then the low energy physics would show a behavior different from QCD. 
The theory is scale invariant but interacting, and the chiral symmetry is preserved in IR region.

In recent studies to search for the infrared fixed point (IRFP), in particular, there are many independent studies in the case of SU($3$) gauge theory coupled with $N_f=12$ fundamental fermions.
The existence of the IRFP in $N_f=12$ theory was predicted by the perturbative beta function at $2$-loop~\cite{Caswell:1974gg} and higher~\cite{Vermaseren:1997fq} in the $\overline{\mathrm{MS}}$ scheme.
The phase structure of $N_f$ expansion was also studied in the paper~\cite{Banks:1981nn}.
Based on these analytical studies, there are several works using lattice simulations to investigate the nonperturbative running coupling constant and the phase structures, although the results have been controversial hitherto. 
In Ref.~\cite{Appelquist:2009ty}, the running coupling constant was computed in the Schr\"{o}dinger functional
(SF) scheme~\cite{Luscher:1991wu,Luscher:1992an,Luscher:1992ny}, and exhibited scale independent behavior in the IR at coupling $g^{2 *}_{SF} \sim 5$.
And studies with the MCRG method~\cite{Hasenfratz:2010fi,Hasenfratz:2011xn}, studies on the phase structure in the finite temperature system~\cite{Deuzeman:2009mh,Miura:2011mc} and the scaling behavior with mass deformed theory~\cite{Appelquist:2011dp,DeGrand:2011cu} show the evidence of the IRFP. 
In the studies of SU($3$) gauge theory with $N_f=7$ and $10$~\cite{Iwasaki:2003de, Hayakawa:2010yn}, they found that theory is in the conformal window which also suggest that $N_f=12$ theory is conformal.
On the other hand, the studies of the mass scaling behavior~\cite{Fodor:2011tu}
and the spectrum of the Dirac operator and the chiral symmetry~\cite{Fodor:2009wk, Jin:2012dw} show the evidence that this theory is not conformal at low energy.
This situation is confusing, since the existence of the fixed point must be scheme independent.
The scheme independence is understood as follows.
Let us consider a relationship between two renormalized coupling constant defined in different renormalization schemes,
$g_1 =F(g_2)$. The beta functions of these two schemes are related as
\beq
\beta(g_1)= \frac{\partial F}{\partial g_2} \beta(g_2).
\eeq
A zero point of the beta function is thus scheme independent except for singular point of the transformations.

One possible reason for the controversial situation could be the underestimate of the discretization errors.
The nonperturbative running coupling constant using lattice simulations can be obtained using the step scaling method.
This method is established by the paper~\cite{Luscher:1991wu} and the nonperturbative running of  the renormalized coupling constant in the continuum limit can be obtained.
One important point which one should bear in mind is that the careful continuum extrapolation and estimation of the systematic uncertainty are important in the low $\beta$ ($\beta \equiv 6/g_0^2$ where $g_0$ is the bare coupling constant) region. 
However, there is no study of the running coupling constant which takes care of the discretization error carefully at least in the case of $N_f=12$.
For example, in the paper~\cite{Appelquist:2009ty}, the constant continuum extrapolation is taken.
That means the discretization effects, which is the renormalization scheme dependent, is neglected.

Another reason may be the bad choice of the value of $\beta$ .
In several previous works, the specific value of $\beta$ is chosen without any reasons.
In the lattice gauge theory with many flavor improved staggered fermion, it was reported that there is a new bulk phase in the strong coupling regime~\cite{Deuzeman:2012ee, Jin:2012dw,Cheng:2011ic}.
Furthermore, the existence of chiral broken phase in the strong coupling limit for the SU(3) theory with $N_f \le 52$ is also reported~\cite{deForcrand:2012vh}.
If the simulation is performed within the bulk phase, it gives an unphysical results because these bulk and chiral broken phases are not connected with the continuum limit with asymptotically free ultraviolet fixed point.
To avoid such phases, the global parameter search and the determination of the parameter region which is obviously connected to the high $\beta$ region is needed.

This work reports a study of the phase structure and the running coupling constant for SU($3$) gauge theories with $N_f=0$ and $12$.
We use the plaquette gauge and the naive staggered fermion actions.
Firstly, we study the phase structure of these theories with both analytical and numerical methods, and then compute numerically the running coupling constant with the twisted Polyakov loop (TPL) scheme in the deconfinement phase. 
The TPL coupling was proposed by de Divitiis {\it et al.}~\cite{deDivitiis:1993hj} for the SU($2$) case, and we extend it to the SU($3$) theory.
This renormalization scheme has no $O(a)$ discretization error, which is of great advantage when we take the continuum limit.
Another advantage of this scheme is the absence of zero mode contributions thanks to the twisted boundary condition~\cite{'tHooft:1979uj, LW:1986}.
This regulates the fermion determinant in the massless limit, which enables simulation with massless fermions.
In this work, we take the continuum limit carefully, and show the existence of the IRFP in the $N_f=12$ theory if we include the systematic uncertainty coming from the continuum extrapolation.

This paper is organized as follows.
We give a definition of SU($3$) TPL renormalized coupling, and the tree level calculation of the TPL coupling  in Sec.~\ref{sec:def-TPL}.
We show the running coupling constant in the case of quenched QCD theory and the scaling behavior of the scheme and the nonperturbative property of the running coupling constant in Sec.~\ref{sec:quenched}. 
We confirm that the renormalized coupling constant in the TPL scheme approaches to a constant if the theory is in the confinement phase.
We discuss how to distinguish such a fake ``fixed point" and the true one in this renormalization scheme.
In Sec.~\ref{sec:vacuum}, we discuss the vacuum structure and the center symmetry of SU($N_c$) gauge theory, in the system coupled with fermions by the semi-classical analysis in the case of $N_f=12$ theory to define the true vacua in this setup.
We also show the numerical results of the phase structure of massive and massless $N_f=12$ SU($3$) theory to determine the parameter region suitable for IRFP search in Sec.~\ref{sec:Nf12-phase-structure}.
In Sec.~\ref{sec:Nf=12}, we study the running coupling constant for massless $N_f=12$ case.
We firstly study the global behavior of the step scaling function using the data set given in Appendix~\ref{sec:app-Japan-data}, and then investigate the existence of the IRFP by the local fit using the additional data set given in Appendix~\ref{sec:app-Taiwan-data} in the strong coupling regime.
The detailed discussion for the stability of the IRFP is given in Appendix~\ref{sec:several-local-fit}.
We also discuss the taste breaking effects in this analysis in Appendix~\ref{sec:Eigen}.

One of the main results of this paper is that we found the IRFP at
\beq
g^{\ast 2}_{\mathrm{TPL}} =2.69 \pm 0.14\, (\mbox{stat.}) _{-0.16}^{+0}\, (\mbox{syst.}),
\eeq
and the critical exponent of the $\beta$ function at the IRFP is
\beq
\gamma_g^\ast = 0.57^{+0.35}_{-0.31} (\mbox{stat.}) _{-0.16}^{+0}\, (\mbox{syst.}).
\eeq

There is an independent paper using the similar idea~\cite{Ogawa-paper}.
In the present work, we add the discussion of the quenched QCD and whose fake fixed point in the TPL scheme.
Also we give a detailed discussion on the vacuum structure and phase structure in $N_f=12$ with the twisted boundary conditions and the taste breaking of the staggered fermion, and add the new data showing the strong evidence of the IRFP beyond the systematic uncertainty.
Data analysis is also refined in various ways as discussed below.
The differences from the paper~\cite{Ogawa-paper}, including the discrepancy of the value of $g^{\ast 2}_{\mathrm{TPL}}$ by more than $2$-$\sigma$, are discussed in detal in Appendix~\ref{sec:app-Taiwan-global} and \ref{sec:app-cont-lim}.
\section{Twisted Polyakov loop (TPL) scheme}\label{sec:def-TPL}
One of the nonperturbative definitions for the renormalized coupling constant can be given by a divergence-free ratio   (${\mathcal A}_{NP}$) of nonperturbative amplitudes.
If the tree level value of the quantity is proportional to the squared bare coupling constant as $ {\mathcal A}_{tree} = k g_0^2$, where $k$ is the constant which is calculated by the tree level quantity, then we can define the nonperturbative renormalized coupling constant from the nonperturbative ratio ${\mathcal A}_{NP}$ by identifying the renormalization factor of the amplitude as the quantum correction of the coupling constant:
\beq
g_{NP}^2 \equiv \frac{ {\mathcal A}_{NP}}{k}.
\eeq
Since the lattice simulation gives us the value of ${\mathcal A}_{NP}$, what we have to do is to find a ratio of tree level amplitudes ${\mathcal A}_{tree}$ which is proportional to the squared bare coupling constant.

Twisted Polyakov loop (TPL) scheme is one of such nonperturbative renormalized coupling schemes defined in finite volume.
This scheme is given in Ref.~\cite{deDivitiis:1993hj} in the case of SU(2) gauge theory, choosing the ratio of Polyakov loop expectation values for twisted and untwisted directions as the quantity ${\mathcal A}_{NP}$.
We extend the definition in Ref.~\cite{deDivitiis:1993hj} to the SU(3) case. 
Although this scheme can be defined in the continuum finite volume, in this section we start a brief review of the definition of TPL scheme on the lattice.
\subsection{The definition of TPL scheme in the SU(3) gauge theory}
To define the TPL scheme, we introduce twisted boundary condition for the link variables ($U_\mu$) in $x$ and $y$ directions  and the ordinary periodic boundary condition in $z$ and $t$ directions on the lattice:
\beq
U_{\mu}(x+\hat{\nu}L/a)=\Omega_{\nu} U_{\mu}(x) \Omega^{\dag}_{\nu},  \label{twisted-bc-gauge}
\eeq
for $\mu=x,y,z,t$ and $\nu=x,y$.
Here, $\Omega_{\nu}$ ($\nu=x,y$) are the twist matrices which have the following properties:
\beq
 \Omega_{\nu} \Omega_{\nu}^{\dag}=\mathbb{I},
 (\Omega_{\nu})^3=\mathbb{I},
 \mbox{Tr}[\Omega_{\nu}]=0, \nonumber
\eeq
and 
\beq
\Omega_{\mu}\Omega_{\nu}=e^{i2\pi/3}\Omega_{\nu}\Omega_{\mu},
\eeq
 for a given $\mu$ and $\nu$($\ne \mu$).
The gauge transformation $U_\mu (r) \rightarrow \Lambda (r) U_\mu (r) \Lambda^\dag (r+\hat{\mu})$ and Eq.~(\ref{twisted-bc-gauge}) imply 
\beq
\Lambda (r+ \hat{\nu}L/a)=\Omega_\nu \Lambda(r) \Omega_\nu^\dag.
\eeq

In the system coupled with fermions, we also have to define the twisted boundary conditions for fermions.
Naively, one might think that the twisted boundary condition for lattice fundamental fermions can be defined by
\beq
\psi (x+\hat{\nu}L/a)=\Omega_{\nu} \psi(x),
\eeq
for $\nu=x,y$.
However, this results in an inconsistency when changing the order of translations, namely,
\beq
\psi (x+\hat{\nu}L/a+\hat{\rho}L/a)&=&\Omega_{\rho} \Omega_{\nu} \psi(x), \nonumber\\
&\ne &\Omega_{\nu} \Omega_{\rho} \psi(x),
\eeq  
for $\rho,\nu=x,y$.
To avoid this difficulty, we introduce a ``smell" degrees of fermion $N_s$~\cite{Parisi:1984cy}, which can be realized by an integral multiple of the number of color symmetry $N_c$. 
We identify the fermion field as a $N_c \times N_s$ matrix ($\psi^a_\alpha$($x$)), where $a$ ($a=1,\cdots,N_c$) and $\alpha$ ($\alpha=1,\cdots,N_s$) denote the indices of the color and smell. 
We can then impose the twisted boundary condition for fermion fields as
\beq
\psi^a_{\alpha} (x+\hat{\nu}L/a)= e^{i \pi/3} \Omega_{\nu}^{ab} \psi^{b}_{\beta} (\Omega_{\nu})^\dag_{\beta \alpha}\label{eq:fermion-bc}
\eeq
for $\nu=x,y$ directions.
Here, the smell index can be considered as a part of  ``flavor'' index, so that the number of flavors should be a multiple of $N_s$, in our case $N_s$ should be the multiple of $N_c=3$.
In our simulation, we use staggered fermion which contains four tastes.
Now, we can label the flavor$(= i, \alpha)$, where $i$ and $\alpha$ denote the taste of staggered and smell indices respectively. 
This restricts the number of flavors to multiples of $12$ in SU(3) gauge theory with twisted boundary condition.

The renormalized coupling in the TPL scheme is defined by taking 
a ratio of Polyakov loop correlators in the twisted ($P_x$) and untwisted ($P_z$) directions:
\beq
g^2_{\mathrm{TPL}}=\lim_{a \rightarrow 0} \frac{1}{k_{latt}} \frac{\langle \sum_{y,z} P_{x} (y,z,L/2a) P_{x} (0,0,0)^{\dag} \rangle}{ \langle \sum_{x,y} P_{z} (x,y,L/2a) P_{z} (0,0,0)^{\dag} \rangle }.\label{TPL-def}
\eeq 
Because of the twisted boundary condition, the definition of Polyakov loops in the twisted directions are modified as,
\beq
P_{x}(y,z,t) ={\mbox{Tr}} \left( [ \prod_{j} U_{x}(x=j,y,z,t)] \Omega_{x} e^{i2\pi y/3L} \right),\label{eq:def-twisted-Poly}
\eeq
in order to satisfy gauge invariance and translational invariance.
At tree level, this ratio of Polyakov loops is proportional to the bare coupling.  
The factor on the lattice ($k_{latt}$) is obtained by analytically calculating the one-gluon-exchange diagram.
In our simulation, we choose the explicit form of the twist matrices~\cite{Trottier:2001vj},
\beq
\Omega_x=\left( 
\begin{array}{ccc}
0 & 1 & 0\\
0 & 0 & 1\\
1 & 0 & 0
\end{array}
\right),
\Omega_y=\left( 
\begin{array}{ccc}
e^{-i2\pi /3} & 0 & 0\\
0 & e^{i2\pi /3} & 0\\
0 & 0 & 1
\end{array}
\right).
\eeq 
The Feynman rule for the SU($N_c$) gauge theory on the lattice with the twisted boundary condition is given in Appendix B in the paper~\cite{deDivitiis:1993hj}.
The value of $k_{latt}$ is given as
\beq
k_{latt}=\frac{1}{g^2 N_c}\frac{1}{\hat{L}^2} \sum_{\hat{k}_\mu} \frac{\exp(i \hat{k}^{ph}\cdot \hat{r})}{\sum_\mu \sin^2(\hat{k}_\mu /2)},
\eeq
where $\hat{L}=L/a$, $\hat{r}=(x,y,z,L/2a)$ and $\hat{k}_\mu$ denotes the momentum in each direction.
In the twisted direction, $\hat{k}_{\mu}$ is given by the sum of  the physical and the unphysical twisted momenta:
\beq
\hat{k}_{x,y} &=& \hat{k}_{x,y}^{ph}+\hat{k}_{x,y}^{\perp},\nonumber\\
&=&\frac{2\pi n_{x,y}^{ph}}{\hat{L}}+ \frac{\pi (2m_{x,y}^{\perp}+1)}{3\hat{L}},\nonumber\\
\hat{k}_{z,t} &=& \hat{k}_{z,t}^{ph},\nonumber\\
&=&\frac{2\pi n_{z,t}^{ph}}{\hat{L}}, \label{eq:k-mu}
\eeq
where $n_\mu^{ph}=0,\cdots \hat{L}/2-1$ and $m_{x,y}^{\perp}=0,1,\cdots, N_c-1$ with $(m_x^\perp,m_y^\perp) \ne (0,0)$.
The momentum $\hat{k}^\perp$ can be identified as the color degree of freedom $(N_c^2-1)$ in the color basis (see the Appendix B in the paper~\cite{deDivitiis:1993hj}).
\begin{table}
\begin{center}
\begin{tabular}{|c|c|}
\hline
$L/a$ & $k_{latt}$\\
\hline
4 &  0.03213022128143844\\
6 &  0.03196454161502177\\
8 &  0.03191145402091543\\
10 &  0.03188777626443608\\
12 &    0.03187515361346823\\
16  &   0.03186277699696222\\
20  &   0.03185710526062057\\
\hline
\end{tabular}
\caption{The value of $k_{latt}$ for $L/a=4,6,8,10,12,16,20$ for SU($3$) gauge theory.}\label{table:k-latt}
\end{center}
\end{table}

In the continuum limit the proportionality factor ($k$) for SU($3$) gauge theory is calculated analytically as:
\beq
k&=&\frac{1}{24 \pi^2} \sum \frac{(-1)^n}{n^2+(1/3)^2}\nonumber\\
&=&\frac{1}{24 \pi^2} \left[ \frac{9}{2}-\frac{3\pi}{2} cosech \left(\frac{\pi}{3} \right) \right]\nonumber\\
&=&0.03184\cdots .
\eeq
The values of $k_{latt}$ in Table~\ref{table:k-latt} can be fitted by a linear function of $O(a^2)$ instead of $O(a)$, as expected.

\section{The TPL coupling for the quenched QCD}\label{sec:quenched}
In this section, we study the TPL coupling constant in the quenched QCD.
In the first subsection, we discuss the property of TPL coupling in the quenched theory and investigate the parameter region where the study of the TPL coupling makes sense.
We obtain the running coupling constant for the quenched QCD in Sec.~\ref{sec:quenched-running}.

\subsection{Phase structure and TPL coupling constant}\label{sec:quenched-phase-str}
The TPL coupling constant is defined by taking the ratio of the correlators of Polyakov loop in the twisted and the untwisted directions.
If the theory is in the confinement phase the correlation length of the Polyakov loop is shorter than the volume, and the gluon does not feel the boundary effect.
In such a situation, we can expect that the ratio of the Polyakov loop correlators becomes unity, and give a fake fixed point.
For this reason, it is awkward to extract the running coupling and try to give a physical meaning to it in such region.
The quenched QCD theory shows the confinement/deconfinement phase transition in the finite volumes, and we can use the TPL running coupling only in the deconfinement phase, where the magnitude of the Polyakov loop shows nonzero values.

To see the property of the TPL coupling in both confined and deconfined phases, we study $\beta$ dependence of the coupling constant at fixed lattice sizes. 
Apart from discretization errors, the coupling increases as $\beta$ decreases at a fixed lattice size.
In this test, we use smaller lattice sizes, $L/a = 2$ -- $6$, with 
relatively low $\beta$ values.
The configurations are generated by the hybrid Monte Carlo algorithm with the Wilson plaquette gauge action.
We measure the Polyakov loop and its correlator for every Monte Carlo trajectory, and each data has the same statistics of $20,000$ trajectories.

\begin{figure}[h]
\begin{center}
  \includegraphics*[height=10cm]{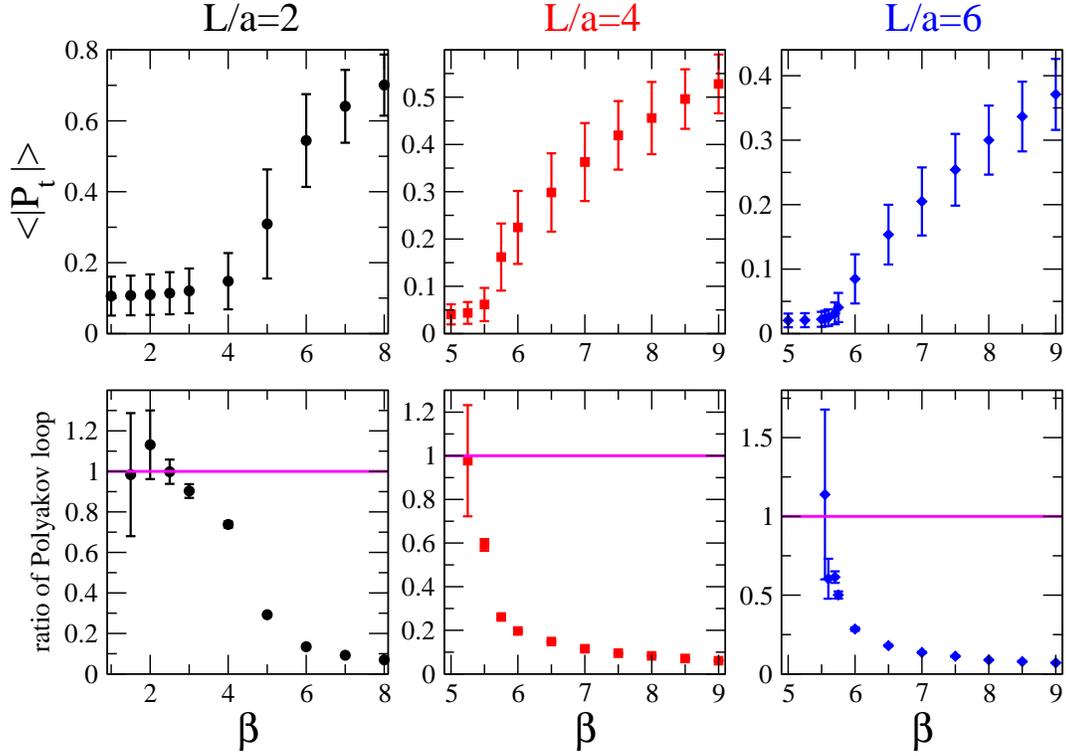} 
\caption{The ratio of Polyakov loop and the absolute value of Polyakov loop in $t$ direction for $L/a=2,4$ and $6$ .}
\label{fig:quenched-FLa}
\end{center}
\end{figure}

The TPL coupling constant and the absolute value of the Polyakov loop in $t$-direction are presented in Fig.~\ref{fig:quenched-FLa}\footnote{We drop the data of the ratio of Polyakov loop for $\beta=1.0,L/a=2$, $\beta=5.0,L/a=4$ and $\beta \le 5.5, L/a=6$, since these error bars become huge. They are consistent with $1$ as expected.}.
The top panels denote the absolute values of the Polyakov loop and the bottom ones denote the corresponding TPL coupling scaled by the coefficient $k_{latt}$ for each lattice size.
We found that the absolute value of the Polyakov loop approaches zero in the low energy region.
The confinement/deconfinement phase transition occurs at the transition point of $\beta$ which depends on the lattice sizes.
From the bottom panels, we can see the ratio of Polyakov loop ($k_{latt} g^2_{\mathrm{TPL}}$) becomes unity below the transition point.

Since there can be a fake fixed point due to confinement, there is a question whether we can use this TPL scheme for the conformal fixed point search in IR region.
One way to judge that the fixed point is not the fake one is to check the the value of renormalized coupling.
Assume that a theory has IRFP. 
The fake fixed point appears at $g^2_{\mathrm{TPL}} \sim 1/k \sim32$.
If there is an IRFP at $g^{2*} _{\mathrm{TPL}} \ne1/k \sim 32$, then we can tell that the fixed point as a physical fixed point.
The other important check is to see the phase structure of the theory at the same time.
At the true conformal fixed point, the theory must be in the deconfinement phase.
There is a possibility of the existence of the bulk phase in the low $\beta$ region, in which the Polyakov loop shows the confinement and/or chiral symmetry breaking~\cite{Deuzeman:2012ee,deForcrand:2012vh,Cheng:2011ic}, if the lattice theory contains the dynamical fermions.
We discuss this point in the case of $N_f=12$ in Sec.~\ref{sec:Nf12-phase-structure}.

\subsection{Running coupling constant for quenched QCD}\label{sec:quenched-running}
\begin{figure}[t!]
\begin{center}
  \includegraphics*[height=7cm]{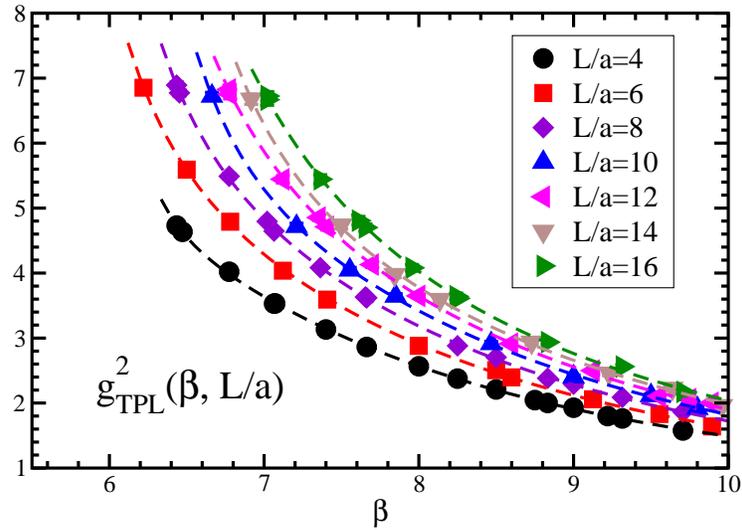} 
\caption{TPL renormalized coupling in the each $\beta$ and $L/a$ 
in quenched QCD. Dashed lines express fit lines of fixed $L/a$ as a function of $\beta$.}
\label{fig:quenched-global-fit}
\end{center}
\end{figure}
Now, we would like to present the results for the running coupling constant in quenched QCD.
The main result was already reported in~\cite{Bilgici:2009nm}.
The gauge configurations in quenched QCD are generated by the pseudo-heatbath algorithm
and overrelaxation algorithm mixed in the ratio $1$:$5$.
One combination of the pseudo-heatbath and $5$ overrelaxation steps called a ``sweep" in the following.
In order to generate the configurations with the twisted boundary condition
we use the trick proposed by L\"uscher and Weisz~\cite{LW:1986}.
To reduce large statistical fluctuation of the TPL coupling constant, as reported
in Ref.~\cite{deDivitiis:1994yz}, 
we measure Polyakov loops at every Monte Carlo sweep and perform 
a jackknife analysis with the bin size of $O(10^3)$.
The simulations are carried out with lattice sizes $L/a=4,6,8,10,12,14,16$
at more than twenty $\beta$ values in the range $6.2 \leq \beta \leq 16$.
We generate 200,000-400,000 sweeps for each parameter set 
$(\beta, L/a)$.

To investigate the evolution of the renormalized
running coupling, we use the step scaling method~\cite{Luscher:1991wu}.
Firstly we choose a value of the renormalized coupling $u{=}g^2_{\mathrm{TPL}}(\beta,a/L)$ at the energy scale $\mu=1/L$.
For each $L/a$ in the set of reference lattice size,
we find the value of $\beta$ which produces a given value of the renormalized coupling, $u$.
Then, we measure the step scaling function on the lattice
\beq
\Sigma (u,a/L;s){=} g^2_{\mathrm{TPL}}(\beta,a/sL)|_{g^2_{\mathrm{TPL}}(\beta,a/L)=u},
\eeq
at the tuned value of $\beta$ for each lattice size $sL/a$.
Here, $s$ is the step-scaling parameter.
The step-scaling function in the continuum limit $\sigma (s,u)$ 
is obtained by taking the continuum extrapolation of $\Sigma (u,a/L;s)$:
\beq
\sigma (s,u)= \lim_{a \rightarrow 0} \Sigma(u, a/L;s)|_{g^2_{\mathrm{TPL}}(\beta,a/L)=u}.
\label{eq:cont}
\eeq
This step scaling function ($\sigma(s,u)$) corresponds to the renormalized coupling at the scale $\mu=1/sL$.
To obtain the running coupling constant in the broad range of scale, we identify the value of $\sigma(s,u)$ with the new input value $u$ at the energy scale $\mu=1/sL$ and repeat the procedures, and then obtain the step scaling function $\sigma(s,u)$ which corresponds to the renormalized coupling at the lower energy scale $\mu=1/s^2L$.  
Repeating this procedure, we can recursively obtain the renormalized couplings at the scales $\mu=1/s^nL$ ($n=0, 1, 2,\cdots $).

Figure~\ref{fig:quenched-global-fit} shows the $\beta$ dependence of the 
coupling constant in TPL scheme at various lattice sizes.
The results are fitted at each fixed lattice size 
to the interpolating function which is similar to the one used in Ref.~\cite{Appelquist:2009ty},
\begin{equation}
g^2_{\mathrm{TPL}}(\beta) = \sum_{i=1}^n \frac{A_i}{(\beta - B)^i},
\end{equation}
where $A_i$ are the fit parameters, and $4\leq B \leq 5$, $n=3,4$ are employed.
As reference lattice sizes of the step scaling, we use $L/a=4,6,8,10$.
The step scaling parameter is $s=1.5$, and we estimate the coupling constant 
for $L/a=9,15$ from interpolations at the fixed $\beta$ using the above
fit results of all the lattice sizes.

\begin{figure}[h]
\begin{center}
  \includegraphics*[height=7cm]{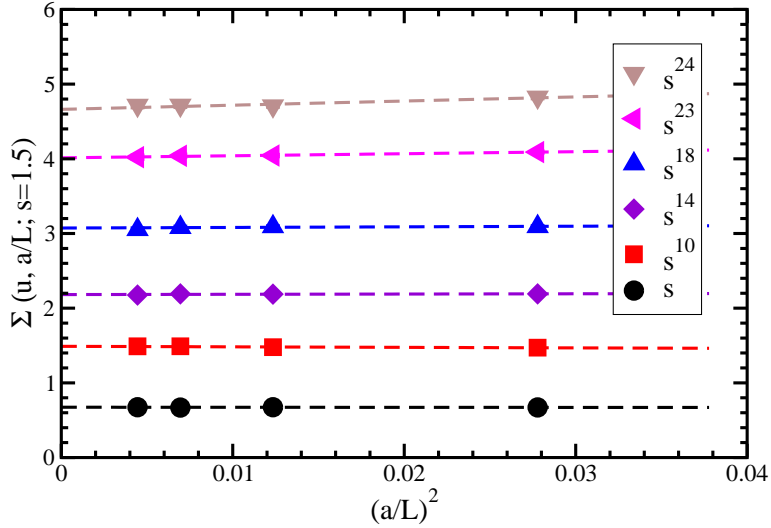} 
\caption{The continuum extrapolations of $g^2_{\mathrm{TPL}}$ for the scale $L/L_0=s^n$. The fit function is a linear 
function in $(a/L)^2$. The data from left to right correspond to
$L/a = 15, 12, 9, 6$.
The statistical error bars are of the same size of the symbols.}
\label{fig:quenched-cont-lim}
\end{center}
\end{figure}

We take the continuum limit using a linear function in $(a/L)^2$, because the 
TPL scheme involves no $O(a/L)$ error.
We found that the coupling constant in the TPL scheme exhibits a nice scaling 
behavior
even at the smaller lattice sizes, as shown in Fig.~\ref{fig:quenched-cont-lim}.

\begin{figure}[h]
\begin{center}
  \includegraphics*[height=7cm]{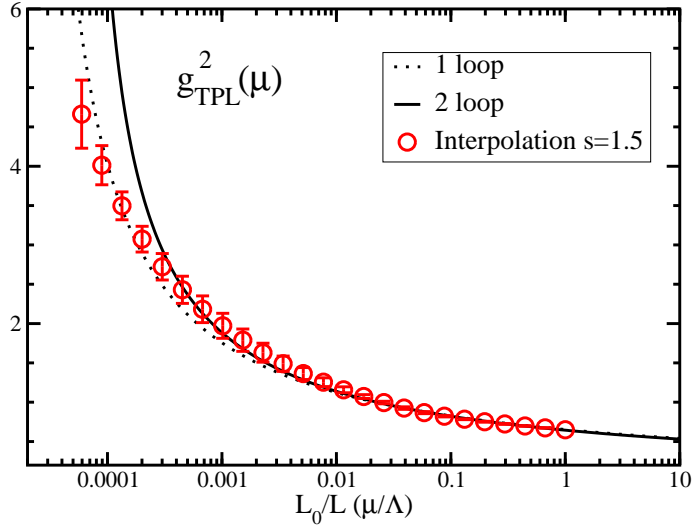} 
\caption{The running coupling constants in TPL scheme, and
in perturbative one-loop and two-loop calculations.}
\label{fig:quenched-running}
\end{center}
\end{figure}

The running of the TPL coupling constant in quenched QCD with 24 steps
is shown in Fig.\ref{fig:quenched-running} together with one- and two-loop perturbative results.
The horizontal axis corresponds to the energy scale.
The energy scales are normalized at $L=L_0$ with $g^2(L_0) = 0.65$.
The nonperturbative running coupling constant is consistent with one- and two-loop perturbative 
results in the high energy region ($L_0/L \geq 0.1$).
Comparison with the other nonperturbative analyses in SF scheme and Wilson loop scheme are also interesting.
Both the  renormalized couplings in SF scheme (Fig.~$1$ in the paper~\cite{Capitani:1998mq}) and Wilson loop scheme (Fig.~$8$ in the paper~\cite{Bilgici:2009kh}) show similar behavior; {\it i.e.,} it runs faster than the one in one-loop result in the nonperturbative region.
On the other hand the TPL running runs slightly slower than the one-loop result. 
  
From this quenched study, we conclude that we can control both the
the statistical and systematic errors of the TPL coupling constant.
Furthermore we find that the TPL coupling constant in quenched QCD
has a robust scaling behavior even in a small lattice size, which was also
observed in the previous quenched SU($2$) calculations~\cite{deDivitiis:1993hj,deDivitiis:1994yz}.

\section{Vacuum structure of the $N_f=12$ theory with twisted boundary condition}\label{sec:vacuum}
Next, we consider the SU(3) theory coupled to $N_f=12$ fundamental fermions. 
In this section, we discuss the center symmetry of SU(3) gauge group to define the true vacuum in this theory.
The generators of the center symmetry of SU($N_c$)
pure gauge theory are $z = \exp(2 \pi i l/N_c)$, where $l = 0, 1, \cdots, N_c-1$.
This symmetry is broken by adding fermions to the theory,
leading to the existence of true vacua in an
SU(3) gauge theory involving massless fermions in the
deconfinement phase.
We discuss the vacuum structure which is important to the study of the gauge theories in finite
volume.

Let us focus on the center symmetry of this theory.
Although the Wilson gauge action is invariant under the following transformation for the link variable for each direction,
\beq
U_\mu(t,\vec{x}) \rightarrow z U_\mu(t,\vec{x}),\label{eq:Z3-transf}
\eeq
the fermion is not invariant.

At the perturbative one-loop level, the semi-classical free energy for the gauge configuration $\{ U \}$ is given as
\beq
F^{\mbox{(tree and one-loop)}} &\equiv& S_g (U) + S^{\mbox{one-loop}}_g[U]  - N_f \ln \det [D(U)].
\eeq
With the twisted boundary condition, the flat potential due to the toron contribution is lifted because of the nonzero momenta in twisted directions and the free energy has $3^4=81-$fold degenerate classical minima at $U_\mu=\exp(2\pi i \theta_\mu/3) \mathbb{I}$, where $\theta_\mu =0,1,2$ for each direction.
The Wilson gauge action ($S_g$) and the one-loop contribution from gauge part ($S_g^{\mbox{one-loop}} [U]$) respects the $Z_3$ symmetry, so that we do not consider them in what follows.

Let us consider the fermion determinant.
In the momentum space, there are the physical and unphysical momenta ($\hat{k}_\mu=\hat{k}^{ph}_\mu+\hat{k}^{\perp}_\mu$) in the twisted directions, that also appear in the gauge field momenta (Eq.~(\ref{eq:k-mu})).
In the case of the fermion field, the color and smell degree of freedom of $\psi_\alpha^a$ in Eq.~(\ref{eq:fermion-bc}) can be transferred into the unphysical momentum degrees of freedom: their number is $N_c \times N_s$
\footnote{In the case of $N_s=n_s \times N_c$, there are $n_s$ flavor fermions whose momentum in the twisted directions have $N_c \times N_c$ unphysical modes.}.
Here, we replace the momentum as
\beq
\hat{k}_\mu \rightarrow \hat{k}_\mu^\theta \equiv \hat{k}_\mu +2\pi i \theta_\mu /3\hat{L}.\label{eq:k-mu-theta}
\eeq
The $Z_3$ transformation in Eq.~(\ref{eq:Z3-transf}) can be defined on each lattice site independently, so that we can take a typical gauge in which $U_\mu= \exp (2\pi i \theta_\mu /3\hat{L})\mathbb{I}$ for whole lattice volume.
Then the fermion action in the vacuum $U_\mu=\exp(2\pi i \theta_\mu /3\hat{L})\mathbb{I}$ is obtained by the above replacement.
The fermion determinant in finite volume $\hat{L}^4$ is thus
\beq
-\ln \det [D] = -8 \sum_{\hat{k}} \ln \left( \sum_{\mu} \sin^2 (\hat{k}_\mu^\theta) \right).\label{eq:fermi-det}
\eeq

In Table~\ref{table:effective-potential}, we give the results of the fermion determinant (Eq.~(\ref{eq:fermi-det})) for $L/a=6$.
\begin{table}
\begin{center}
\begin{tabular}{|c|c|}
\hline
($\theta_z,\theta_t$) & $-\ln \det (D) -S_0$ \\
\hline \hline
($0,0$) & 0 \\
\hline
($0,1 \mbox{ or } 2$) or ($1\mbox{ or }2$,0) & -57.19 \\
\hline
($1 \mbox{ or } 2,1 \mbox{ or } 2$) & -89.56\\
\hline
\end{tabular}
\caption{The $\theta_{z,t}$ dependence of semi-classical free energy for $6^4$ lattice. We take a reference potential energy $S_0=-3511.68$ which is the one in $(\theta_z,\theta_t)=(0,0)$.}\label{table:effective-potential}
\end{center}
\end{table}
We find that the vacuum free energy is independent of $\theta_{x,y}$ and there are three types of vacua classified with $\theta_{z,t}$.
The first one is a ``trivial vacuum", in which vacuum $(\theta_{z},\theta_{t})=(0,0)$. This vacuum has $9-$fold degeneracies.
The value of the free energy is highest among three types of vacuum, so that it will decay to the true vacuum.
The second one is a ``half-trivial vacuum", in which one of $\theta_{z,t}$ is $1$ or $2$ and the other one is $\theta_\mu=0$.
This vacuum has $36-$fold degeneracies. 
The free energy is higher than the one of the third vacuum, so that this vacuum is also unstable.
The third one, in which the free energy is lowest, is a ``non-trivial vacuum".
Both $\theta_{z}$ and $\theta_t$ take $1$ or $2$, and there are also $36-$fold degeneracies.
The free energy has minima at this vacuum where all classical link variables for $z$ and $t$ directions has a non-trivial phase $U_{z,t} \propto \exp(\pm 2\pi i/3\hat{L})$.
This means that the Polyakov loop for $z$ direction also has a non-trivial phase $\exp(\pm 2\pi i/3)$.

This classification holds for generic lattice size, and it turns out that the difference of the free energy between the true non-trivial vacua and the other vacua becomes small as the lattice size becomes larger, where as the potential barrier becomes higher.
If we change the fermion boundary condition in $z$ and $t$ directions, the semi-classical free energy has minima at other vacua. 

\section{Simulation setup for the $N_f=12$ theory}
Our numerical simulation is performed in the following setup.
The gauge configurations are generated by the hybrid Monte Carlo 
algorithm with the Wilson gauge and the naive staggered fermion actions.
In Sec.~\ref{sec:Nf12-phase-structure}, the simulations are carried out with lattice sizes $L/a=4,8$ and $12$
at several low $\beta$ and a broad range of $ma$ to study the phase structure in this system.
We generate $1,000$ -- $2,000$ trajectories for each parameter set in the case of $L/a=4$ and $8$,
and also generate $500$ -- $1,000$ trajectories for each in the case of $L/a=12$.

The measurement of the coupling constant in Sec.~\ref{sec:Nf=12} are carried out with lattice sizes $L/a=6,8,10,12,16$ and $20$
\footnote{We generated the data with $L/a=4$ as same as the quenched QCD case, however, we found there are large discretization effects in the case of $N_f=12$~\cite{Itou:2010we}, and the systematic uncertainty could not be controlled. Therefore we dropped the data from this analysis.}
at around thirty values of $\beta$ in the range $4.0 \leq \beta \leq 100$ with $ma=0$.
To reduce statistical fluctuations, we measure the Polyakov loops at every trajectory and bin the data by taking the autocorrelation into account.
Using the jackknife method, typical statistical errors of correlator are found to be $2$ -- $3 \%$.
We also estimate the statistical error within the bootstrap method in the whole analysis in Sec.~\ref{sec:Nf=12}, and obtained consistent results.

\section{Simulation results: Phase structure of $N_f=12$ SU($3$) theory with the twisted boundary condition}\label{sec:Nf12-phase-structure}
In this section, we investigate the phase structure of the $N_f=12$ fermion theory on the lattice.
Although the main purpose of this paper is to study the running coupling in the massless theory, in order to fully understand the phase structure we need to understand the phase structure of the whole region of the theory space including the mass parameter.
Actually, there are several studies which reported the existence and absence of the bulk phases in the case of $N_f=12$ staggered fermion system~\cite{Cheng:2011ic,Deuzeman:2012ee}.
In the paper~\cite{Cheng:2011ic}, the authors found the existence of a bulk phase where shift symmetry and chiral symmetry are weakly broken, and the paper~\cite{Deuzeman:2012ee} reported that it is caused by the improvement of the staggered fermion.
Furthermore, there is the spontaneous chiral symmetry broken phase in the strong coupling limit for $N_f \le 52$~\cite{deForcrand:2012vh, Tomboulis:2012nr} for SU(3) massless fermion theory.
In our simulation, we use the naive staggered fermion and introduce the twisted boundary condition.
It is important to show the phase structure in our lattice setup independently to identify the parameter range suitable for the study the TPL coupling constant.

In Sec.~\ref{sec:mass-phase} and Sec.~\ref{sec:phase_ploop}, we study the $\beta$ and volume dependence of the expectation value of the plaquette and the Polyakov loops from which we determine the phase structure of the theory in the $\beta-ma$ plane. 
In Sec.~\ref{sec:tunneling}, we discuss a contribution of the vacuum tunneling due to the lattice artifact to the TPL coupling constant.
Due to the finite lattice spacing, the tunneling behavior between these vacua occurs during Monte Carlo simulation and it must be a lattice artifact.
As we have shown using semi-classical analysis in Sec.~\ref{sec:vacuum}, there are the degenerate vacua in this lattice setup.  
We show that  the contribution is negligible in our simulation.

\subsection{Plaquette values on the $\beta-ma$ plane}\label{sec:mass-phase}
\begin{figure}[h]
\begin{center}
   \includegraphics*[height=9cm]{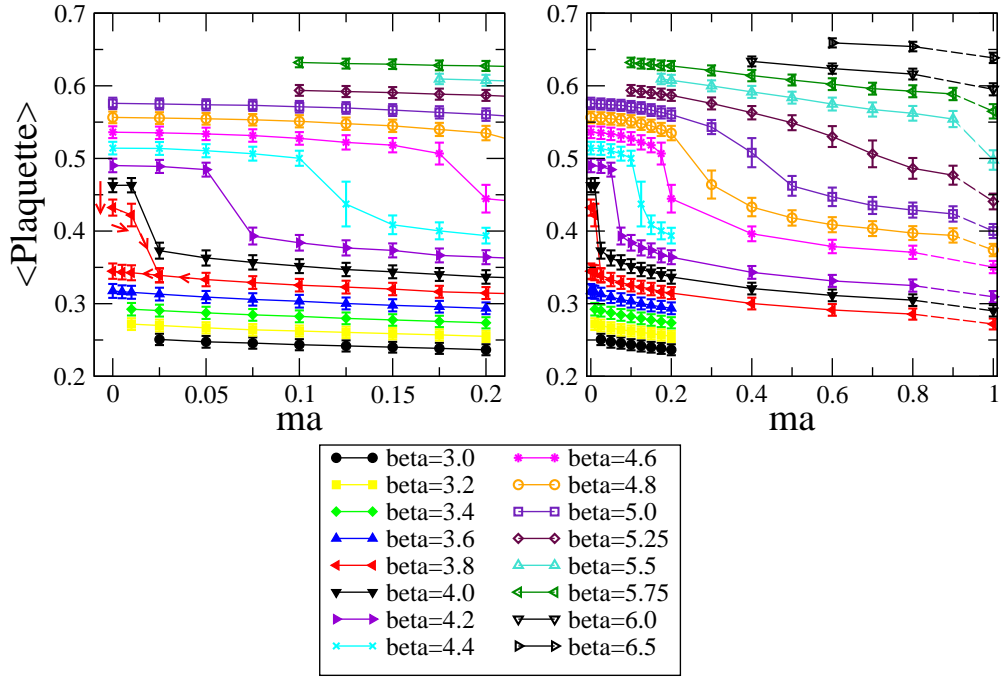}
  \end{center}
\caption{Plaquette values for several $\beta$ and $ma$ on $(L/a)^4=4^4$ lattice. The data at $ma=1$ on the right panel denotes the quenched QCD. The small (red) arrows near the massless at $\beta=3.8$ on the left panel shows the detailed history of the thermalization. The other data are thermalized from massive to massless direction.}
\label{fig:Plaq-L-4}
\end{figure}

Let us investigate the plaquette values of the $\beta$--$ma$ plane.
The left panel of figure~\ref{fig:Plaq-L-4}  shows the plaquette values on $(L/a)^4=4^4$ lattice in the $\beta$--$ma$ plane in the range of $0 \le ma \le 0.2$.
Most of the configurations are thermalized from massive to massless direction except for the small mass region in the $\beta=3.8$ as we explain later. 

First of all, we found there are discontinuities in the plaquette as a function of $\beta$ along the lines of fixed $ma$.
The discontinuity appears between $\beta=4.8$ and $\beta=4.6$ for $ma=0.2$ and in the smaller mass region it appears at the lower $\beta$.
Near the massless limit the gap exists around $3.6 \le \beta \le 4.0$.

The small (red) arrows near the massless at $\beta=3.8$ on the left panel in Fig.~\ref{fig:Plaq-L-4} shows the detailed histories of the thermalization.
We find that there are two different values in the $0 \le am \le 0.0125$ region.
The configurations giving larger values of the plaquette at the same $ma$ are generated starting from configuration with massless fermions in $\beta=4.0$; on the other hand those giving smaller values are obtained starting from the configuration with massive fermions at fixed $\beta$.
The hysterisis clearly indicates that there is a first order phase transition around this region.
At $\beta=4.0$ and $\beta=3.6$, there is no dependence on the thermalization process.

We also study larger mass region.
The right panel in Fig.~\ref{fig:Plaq-L-4} is the same plot as the left one for a broader region of $ma$.
In the quenched limit, we know that there is the first order phase transition.
In the figure, we plot the data for the quenched lattice at $ma=1.0$.
The gap seems milder in the larger mass region.

\begin{figure}[h]
\vspace{0.3cm}
\begin{center}
   \includegraphics[height=10cm]{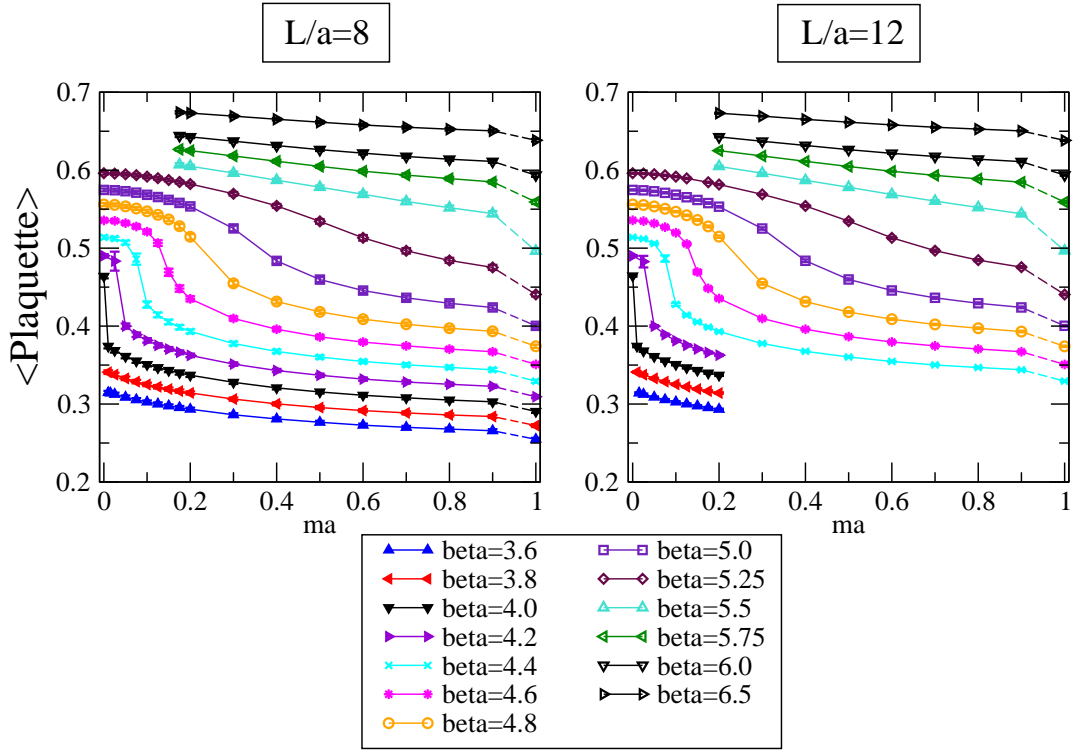}
  \end{center}
\caption{Plaquette values for several $\beta$ and $ma$ on $(L/a)^4=8^4,12^4$ lattice. The data at $ma=1$ denotes quenched QCD. The statistical error is the same size of the symbol.}
\label{fig:Plaq-L-8-from-quenched}
\end{figure}
We also investigate this first order phase transition near the massless region by changing the lattice volume in Fig.~\ref{fig:Plaq-L-8-from-quenched}.
There are slight differences between $L/a=4$ and $L/a=8$ for the critical value of $\beta$.
For example at $ma=0.175$ the data for $\beta=4.6$ on $(L/a)^4=4^4$ is in the upper phase of the gap (See Fig.~\ref{fig:Plaq-L-4}), but it moves to the lower phase in the case of $(L/a)^4=8^4$ (See left panel in Fig.~\ref{fig:Plaq-L-8-from-quenched}).
The results on $(L/a)^4=8^4$ and $12^4$ show the similar behavior at least the present interval of $\beta$ and $ma$ ($\Delta \beta=0.2,\Delta ma =0.01$ -- $0.025$), and there is no clear volume dependence.
Since the massless simulation at $\beta=3.8$ needs extremely finer molecular-dynamics time step size than $\Delta \tau =0.002$ ($\tau=1$ is $1$ trajectory), practically we could not generate the data.
The position of $\beta$ where the simulation becomes quite costly is the same for both $L/a=8$ and $L/a=12$.
It suggests that near the massless region there is a bulk phase transition in $\beta < 4.0$.

\subsection{Polyakov loop}  \label{sec:phase_ploop} 
Next, let us investigate the Polyakov loop.
Since the dynamical fermions breaks the center symmetry explicitly, there is no clear order parameter for the deconfinement phase transition. 
However, here we use the word ``deconfinement" phase for the region in the theory space where magnitude of Polyakov loop is clearly nonzero on the lattice.

In the case of the quenched QCD, there is $Z_3$ symmetry, and there is tunneling between them on finite lattices.
In the case of $N_f=12$ massless theory with the twisted boundary condition, the $Z_3$ symmetry is broken and 
the true vacua is the one that the Polyakov loops in the untwisted directions have the nontrivial phase as explained in Sec.~\ref{sec:vacuum}.

\begin{figure}[h]
 \begin{center}
 \vspace{0.3cm}
   \includegraphics[height=7cm]{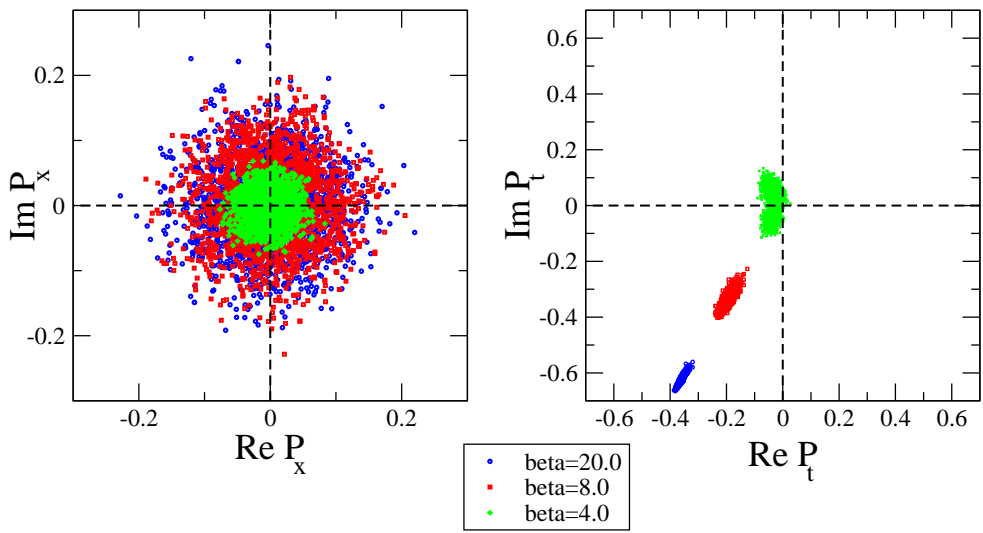}
\end{center}
\caption{Scatter plots of the Polyakov loop in the twisted ($\mu=x$) and the untwisted ($\mu=t$) directions in the complex plane. The lattice size is $6^4$ and circle(blue), square(red) and cross(green) symbols denote $\beta=20.0, 8.0, 4.0$ respectively.}
\label{fig:phase-ploop-P1}
\end{figure}

At first, we present the scatter plots of the typical Polyakov loops for the massless theory in the twisted
and untwisted directions in the left and right panels in Fig.~\ref{fig:phase-ploop-P1} respectively.
The complex phase of the Polyakov loop in the twisted direction does not favor any particular value.
This is consistent with the tree level analysis, in which $ \langle P_x \rangle =0$ because of the traceless twist matrix in the Eq.~(\ref{eq:def-twisted-Poly}).
Since the Polyakov loop expectation value in the twisted direction is zero and does not depend on the value of $\beta$, it is not related to whether the system is in 
the confinement or in the deconfinement phase.
On the other hand, the Polyakov loop expectation value in the untwisted direction has clearly the nontrivial complex phase $\exp(\pm 2\pi i/3)$ in the high $\beta$ region.
In $\beta = 4.0$ the tunneling occurs between the two complex phases,
but apart from the tunneling 
the values of the phase are close to $\exp(\pm 2\pi i/3)$.
Thus, we confirm that the Polyakov loops in both directions are consistent with the results of from the semi-classical
analysis in Sec.~\ref{sec:vacuum} even in the strong coupling region.
The effect of the tunneling on the TPL coupling 
will be discussed in the next subsection.

\begin{figure}[h]
\begin{center}
   \includegraphics[height=7cm]{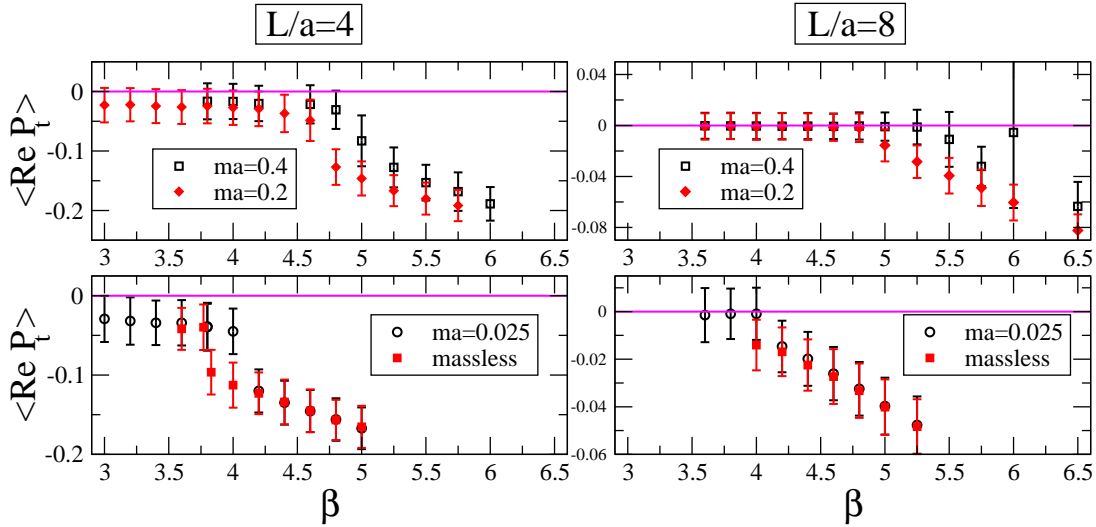}
  \end{center}
\caption{Left panel: real part of the Polyakov loop in $t$ direction in the case of $(L/a)^4=4^4$. Clearly there is a gap for each $ma$. In the case of the massless fermions, the measured data at $\beta=3.8$ are shifted;  one of them shows at $\beta=3.77$ which corresponds to the configurations at $\beta=3.8, ma=0$ whose plaquette value is the larger one in Fig.~\ref{fig:Plaq-L-4}. The other shows at $\beta=3.83$ which corresponds to the configurations at the same $\beta$ and $ma$ whose plaquette value is the smaller one in Fig.~\ref{fig:Plaq-L-4}.
Right panel: the same plot in the case of $(L/a)^4=8^4$. }
\label{fig:RePloop-each-m-L-4}
\end{figure}
Next, we show the real part of Polaykov loop in $t$-direction.
The left two panels in Fig.~\ref{fig:RePloop-each-m-L-4} show the ones at several fixed $ma$ and $\beta$ in the case of $L/a=4$.
We can find that there is a gap of the real part of the Polyakov loop at fixed $ma$ data, and the value of $\beta$ at the gap corresponds to the critical value of $\beta$ of confinement/ deconfinement phase transition.
In the case of massless fermions on $L/a=4$ we find a gap at the $\beta=3.8$.
For $\beta$ smaller than the gap position the real part of the Polyakov loop is not consistent with zero, but it goes to zero continuously.
In the finite mass region, there is a weak jump, and the gap become larger in the smaller mass region.
The value of the critical $\beta$ in which the data shows the jump is the same with the plaquette study.

\begin{figure}[h]
\begin{center}
   \includegraphics[height=6cm]{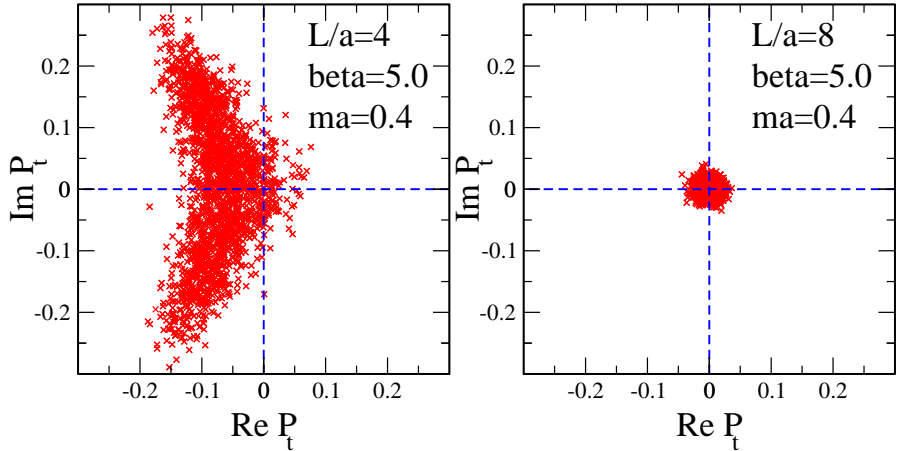}
  \end{center}
\caption{Left panel: real part of the Polyakov loop in $t$ direction in the case of $\beta=5.0, ma=0.4, (L/a)^4=4^4$. Right panel: the same plot in the case of the same $\beta$ and $ma$ on $(L/a)^4=8^4$. }
\label{fig:L-dependence-Phase}
\end{figure}
Let us study the lattice size dependence.
We show the real part of the Polyakov loop in the case of $L/a=8$ in the right panels in Fig.~\ref{fig:RePloop-each-m-L-4}.
There is no clear jump in the case of $L/a=8$, but the real part of the Polyakov loop approach to the zero in the low $\beta$ region.
We define the critical value of $\beta$ as the largest value of $\beta$ for which the real part of Polyakov loop becomes consistent with zero. 
Again, the critical value of $\beta$ is the same with the plaquette study.
We can conclude that in Figs.~\ref{fig:Plaq-L-4} and \ref{fig:Plaq-L-8-from-quenched}, the phase for $\beta$ larger than the gap position can be identified as the deconfinement phase and that for $\beta$ smaller than the gap position is the confinement phase.

Note that there is one misleading exceptional data at $\beta=6.0, ma=0.4, L/a=8$.
The real part of Polyakov loop is consistent with zero. 
However, we can find that the theory is in the deconfinement phase from the scatter plot of the Polyakov loop on the complex plane.
Since at the parameter the fermion mass is too heavy, there is the tunneling between $Z_3$ vacua as in the quenched case.
That is the lattice artifact, so that the data does not say the theory is in the confinement phase.

From the comparison of the data at $L/a=4$ and $L/a=8$, we found there is a volume dependence of the critical value of $\beta$ in the massive region, while there is no dependence near massless region.
Actually, in the top panels of Figs.~\ref{fig:RePloop-each-m-L-4}, the theory with $ma=0.4, L/a=8$ is in the confinement phase for $\beta \le 5.0$, while in the case of $L/a=4$ the theory is in the deconfinement phase for $4.8 \le \beta$.
Figure~\ref{fig:L-dependence-Phase} shows the scatter plots of the Polyakov loop on the complex plane.
These plots show that there is a clear volume difference.
On the other hand, in the case of the (nearly) massless fermion, the transition point does not show the volume dependence, indicating that the transition is bulk one.
The theory with the massless fermion in both $L/a=4$ and $8$ lies in the deconfinement phase for $\beta \ge 4.0$.
At least the current interval of $\beta$ and $ma$, we cannot find the volume dependence in the small mass region.
Since in the quenched limit we know that there is a finite volume phase transition, we expect that there is a crossover for both bulk and the finite volume phase transition in the middle range of the fermion masses.

Furthermore, in the case of $L/a=12$, we cannot find clear nonzero value of the Polyakov loop in the whole region since the current preliminary statistics is small and the Polyakov loop in the low $\beta$ region is noisy.
The gap of the plaquette and the confinement/deconfinement phase transition seems to occur simultaneously, and there is no difference between $L/a=8$ and $L/a=12$ on the plaquette behavior. 
At least we found that the position of $\beta$ where the simulation becomes quite heavy is the same in the case of $L/a=8$.

Finally, we study the phase structure of the massless fermion $N_f=12$ QCD for $\beta \ge 4.0$ and $L/a=6$--$20$.
We find that all configurations, which are used for  the running coupling constant study in Sec.~\ref{sec:Nf=12}, live in the deconfinement phase (although it might be trivial since the transition seems to be the bulk and we concentrate on the parameter region within $\beta \ge 4.0$). 
\begin{figure}[h]
\begin{center}
\vspace{0.2cm}
   \includegraphics[height=6cm]{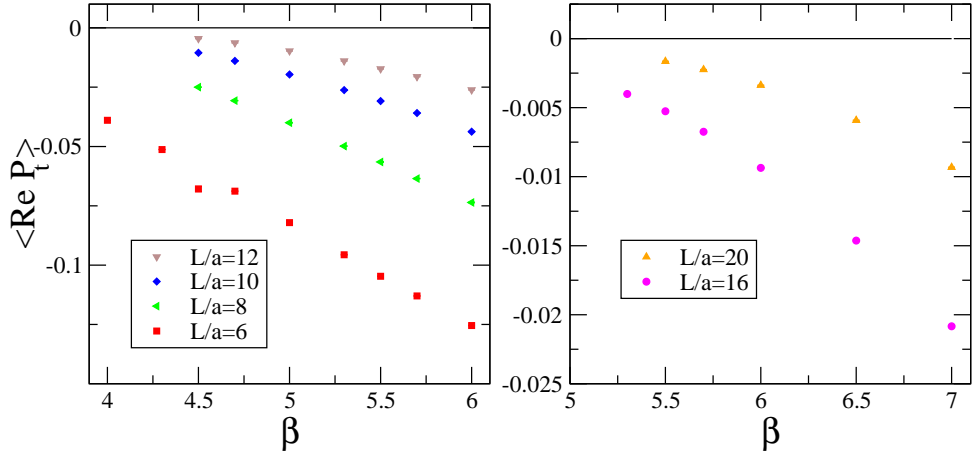}
  \end{center}
\caption{Real part of the Polyakov loop in the untwisted ($\mu=t$) direction
as a function of the lattice size at $\beta = 6.0$. The statistical error bars are the same size of the symbols. }
\label{fig:P4-L-dependence}
\end{figure}
Figure~\ref{fig:P4-L-dependence} shows the real part of the Polyakov loop for the low $\beta$ region for each lattice size.
It can be clearly seen that $\mbox{Re P}_t$ takes nonzero values for the entire data, indicating that all configurations used in our analysis in Sec.~\ref{sec:Nf=12} are in the deconfinement phase.

\begin{figure}[h]
\begin{center}
  \includegraphics*[height=6cm]{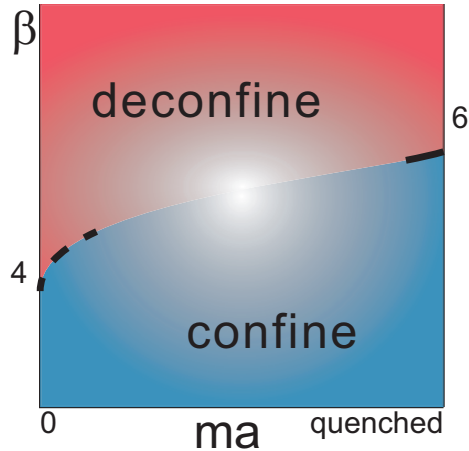} 
\caption{The phase structure of $N_f=12$ SU($3$) theory with naive staggered fermion. The dashed line denotes bulk phase transition and the solid line denotes the finite volume phase transition.}
\label{fig:phase-Nf12}
\end{center}
\end{figure}
The summary of the phase structure and the available region of the TPL coupling for the quenched and the massless $N_f=12$ QCD is the following.
Figure~\ref{fig:phase-Nf12} is a sketch of the phase structure for the naive staggered $N_f=12$ SU($3$) theory.
In the case of the quenched QCD, the correlation length becomes shorter in the lower $\beta$ region, and there is the finite volume phase transition where the theory goes to the confinement phase.
In the case of the massless $N_f=12$ SU($3$) theory, there is the similar behavior while the transition seems to be the bulk one at $\beta < 4.0$.
In the study on the running coupling constant in TPL scheme, we should focus on only the deconfinement phase on the lattice.

Furthermore, in $\beta \ge 4.0$ region with massless fermions, we also investigate the eigenvalue of Dirac operator which is presented in Appendix~\ref{sec:Eigen}.
The lowest eigenvalues are clearly nonzero even in the lowest $\beta$ for all lattice sizes.
According to the Banks-Casher relation, it implies that the chiral symmetry is restored in $\beta \ge 4.0$.
In Sec.~\ref{sec:Nf=12}, we finally find the IRFP at higher $\beta$ values than the bulk phase transition point, although the values of physical critical $\beta$ at physical IRFP depend on the lattice sizes.
Our phase diagram (Fig.~\ref{fig:phase-Nf12}) is completely consistent with the conjectured phase diagram in the paper~\cite{deForcrand:2012vh} (Fig.~{10}) by Ph.~de~Forcrand {\it et al.}, where they study the strong coupling limit.

\subsection{Tunneling behavior between true vacua}\label{sec:tunneling}
During the Monte Carlo simulation, the tunneling can occur 
between the two degenerate vacua in each untwisted direction independently.
The tunneling behavior is a lattice artifact, since the potential barrier between the vacua 
becomes finite at the finite lattice spacing.
In this subsection, we consider how the TPL coupling is disturbed by the tunneling behavior.

The tunneling is expected to occur more frequently in the strong coupling region.
We observe some tunnelings in low $\beta$ region,
although the number of the tunneling is quite small,
and decreases as $\beta$ increases.
For $L/a = 6$ we observe the tunneling $7$ times per $6,000$ trajectories at $\beta=4.0$, 
and only $3$ times per $90,000$ trajectories at $\beta=6.0$.

\begin{figure}[h]
\begin{center}
   \includegraphics*[height=7cm]{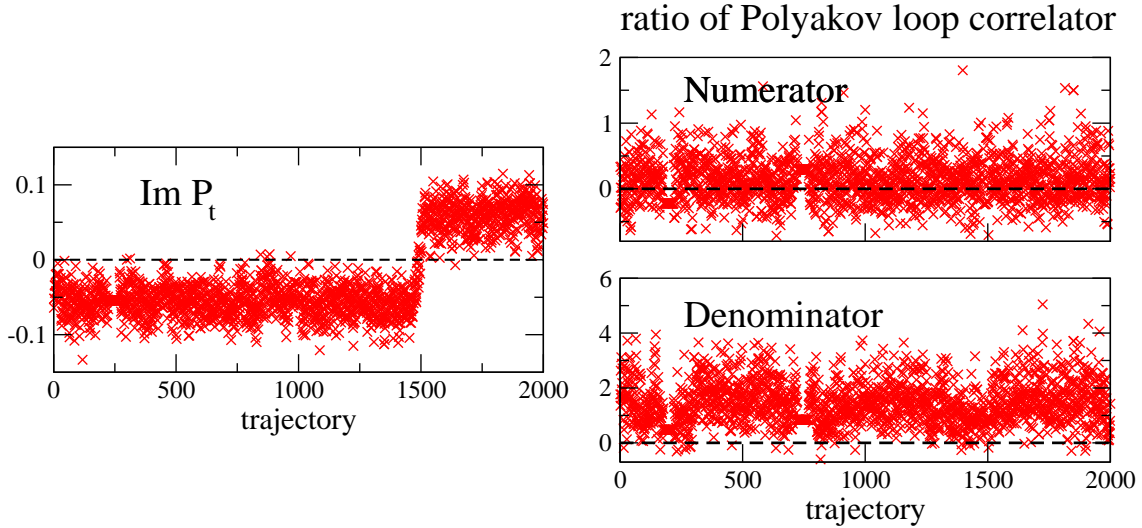}
  \end{center}
\caption{Left panel: History of the imaginary part of the Polyakov loop 
in untwisted ($\mu=t$) direction at $\beta=4.0$ in $L/a=6$.
Right panels: Histories of the numerator and denominator of the TPL coupling
Eq.~(\ref{TPL-def}) in the same trajectories of the left.
}
\label{fig:rPoly-for-each}\label{fig:P4-each-sweep}
\end{figure}

A typical example of the tunneling is shown in 
figure~\ref{fig:P4-each-sweep}.
The left panel denotes the history of the imaginary part of 
the Polyakov loop in the untwisted direction at $\beta=4.0$ and $L/a=6$.
As can be seen in Fig.~\ref{fig:P4-each-sweep}, the sign of the imaginary part is flipped at around the $1,400$ -- $1,600$th trajectories.

Let us see the effects of the tunneling on the TPL coupling defined by the ratio of the Polyakov loop correlators
in Eq.~(\ref{TPL-def}).
The right panels in Fig.~\ref{fig:rPoly-for-each} present the histories of the 
numerator and denominator of the TPL coupling 
obtained from the same configuration 
as in the left panel.
During the tunneling, {\it i.e.,} the $1,400$ -- $1,600$th trajectories,
both the results seem not to exceed the range of each statistical fluctuation.
This means that the tunneling does not significantly affect 
the result of the TPL coupling calculated by the ratio of the
expectation values for the numerator and denominator.

Therefore we conclude that effects of the tunneling on the TPL coupling is negligible at least in our calculation
due to the rare occurrence of the tunneling and the large statistical fluctuation of the Polyakov loop correlators.

\section{Simulation results: Step scaling function for $N_f=12$ SU(3) gauge theory}\label{sec:Nf=12}
In this section, the step scaling function in the case of $N_f=12$ flavor is derived.
As in the quenched QCD case, we use the step scaling method to find the IRFP.
In this section, we focus on another quantity, which is called the growth rate of step scaling function, rather than the running coupling constant.
As we explained in Sec.~\ref{sec:quenched-running}, the running coupling constant is derived using ``recursively" the step scaling procedure. 
Since the error of $\sigma(s,u)$ feeds in to the input renormalized coupling ($u$) in the next step scaling procedure, the errors from $\sigma(s,u)$ accumulate  in the running coupling towards lower energy.
On the other hand, the step scaling function $\sigma(s,u)$ with no error accumulation can be defined independently for each step and the growth rate $\sigma(s,u)/u$ becomes unity when there is a zero in the beta function.
Therefore the growth rate $\sigma(s,u)/u$ is a suitable quantity for searching the IRFP.

At first, we will discuss the global behavior of the growth rate from the perturbative to the IR region in Sec.~\ref{sec:Nf=12-global-fit}.
The nonperturbative running behavior shows the signal of the conformal fixed point in the IR region.
Then, we focus on the low energy region only and derive again the step scaling function by using the data only in the strong coupling region in Sec.~\ref{sec:Nf=12-local-fit}.
We discuss the stability of the IR fixed point by considering several systematic uncertainties and derive the universal quantity for the exponent of the beta function around the IRFP in Sec.~\ref{sec:critical-exp}.
%


\subsection{The global behavior of the step scaling function}\label{sec:Nf=12-global-fit}
Before explaining the simulation details, we address the guiding 
principles of our simulation to show the global behavior of the running coupling.
We use the step scaling method as in the quenched case.
For each $L/a$ we interpolate the data in $\beta$. 
Since the choice of the interpolation function can affect the step scaling function,
we should take care of the following two points:
\begin{enumerate}
\item We generate data in a broad range of $\beta$, with intervals such that the renormalized coupling constant ($g_R^2$) grows
         almost constantly in each interval. Thus the interval of $\beta$ is large in high $\beta$ region 
         while small in the low $\beta$ region. Each data has a similar accuracy ($2 -3\%$). 
\item We employ fit functions for $\beta$ interpolation which reproduce the tree level result 
          $g_R^2 \simeq g_0^2$ on each lattice size in extremely high $\beta$ region. 
\end{enumerate}
These guiding principles ensures the stability of our fit results under the change of 
fit functions and the number of data.
Point $1$ is needed to ensure that the fit functions do not favor any special region 
of the data when we interpolate our data in $\beta$ or extrapolate to in $(a/L)^2$.
To satisfy this point is very important to search for the IRFP by using the numerical simulations,
 since finally the interpolation function of the data decides the step scaling function. 
We discuss the dependence of the data set for the global fit analysis in Appendix~\ref{sec:app-Taiwan-global}, in the case where the data are concentrated in some particular region. 
The result in Appendix~\ref{sec:app-Taiwan-global} do not have a nice agreement with the perturbative result and in the IR region the position of the IRFP strongly depends on the fit range.
Point $2$ is needed to reduce the effect of statistical fluctuation in high $\beta$ region.
In high energy region, since the coupling runs very slowly, we need extreme high statistics to reproduce the perturbative result only by the numerical data.
The assumption of the point $2$ makes the result stable in high energy region.

\begin{figure}[tb]
  \begin{center}
   \includegraphics[height=7cm]{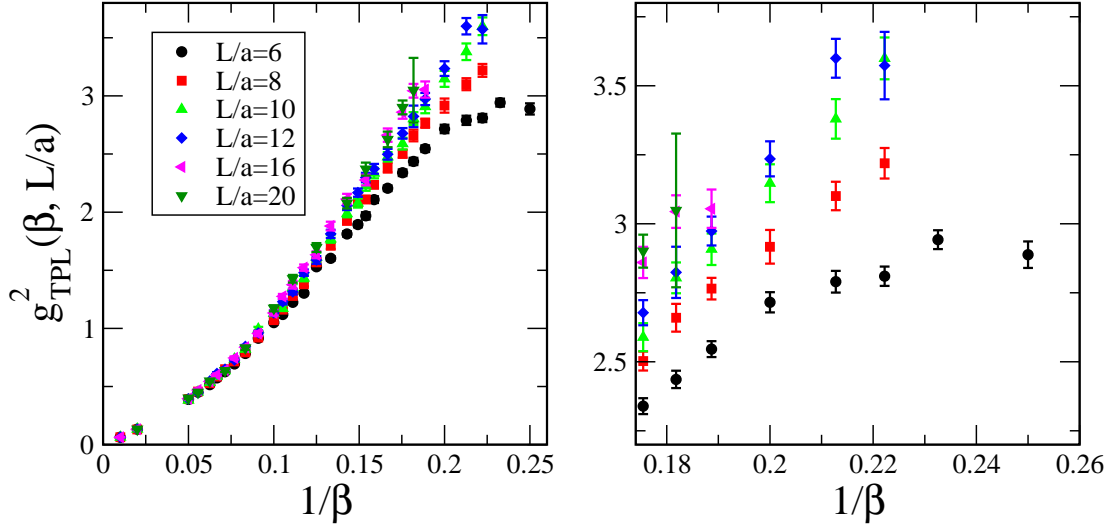}
  \end{center}
  \caption{TPL coupling for each $\beta$ and $L/a$. Left panel: Plots for the global region of $\beta$. Right panel: Plots only for the low $\beta$ region.} 
  \label{fig:global-g2}
\end{figure}

In Fig.\ref{fig:global-g2}, we show our simulation results for 
the renormalized coupling in TPL scheme as a function of $1/\beta$ 
for each $L/a$.
The raw data are given in Tabs.~\ref{tab:g2L6-8-10} and \ref{tab:g2L12-16-20} in Appendix~\ref{sec:app-Japan-data}.
The left panel in the Fig.~\ref{fig:global-g2} shows a global behavior of the TPL coupling.
We can see the high energy behavior seems to be almost linear in $1/\beta$ as expected from the perturbation theory.
The right panel focuses on the low $\beta$ region.
In the low $\beta$ region for $L/a=6$ the TPL coupling has a maximum at $\beta=4.3$.
In contrast to the Schr\"{o}dinger functional scheme~\cite{Appelquist:2009ty}, the renormalized coupling gets larger for larger volume for all the range of $\beta$.
We consider that this difference comes from the lattice artifact which depends on the renormalization scheme.
To remove the effect, the careful continuum extrapolation is necessary.

For $\beta$-interpolation, 
we use the following form of fitting function:
\begin{equation}
g^2_{\mathrm{TPL}}(\beta,a/L) = 6/ \beta
+\textstyle{\sum_{i=1}^{N} C_i (a/L)/ \beta^{i+1}},\label{eq:beta-fit}
\end{equation}
where $C_i (a/L)$ are the fitting parameters and $N$ is the degree of the polynomial.
Here, $N=3$ -- $5$ are employed. 
We drop the data at $\beta=4.0, L/a=6$ from the fit to avoid the double solutions when we solve the $\beta$ values to reproduce the input renormalized coupling ($u$).
This limits us to study the step scaling in the range $u \le 2.94$, where $u=2.94$ is the renormalized coupling constant $g^2_{\mathrm{TPL}}$ at $\beta=4.3,L/a=6$, in this lattice set up.

To investigate the evolution of the renormalized
running coupling, we use the step scaling method as explained in Sec.~\ref{sec:quenched-running}.
In this study, we take $s=1.5$, and denote 
$\sigma(u)\equiv\sigma(s{=}1.5, u)$ in the rest of 
this paper for simplicity. 
The set of small lattices is taken to be 
$L/a=6, 8, 10,12$, therefore,  we need values of 
$g^2_{\mathrm{TPL}}$ for $L/a=9, 12, 15,18$ 
to take the continuum limit in Eq.~(\ref{eq:cont}).
For $L/a=9, 15$ and $18$,  we estimate values of 
$g^2_{\mathrm{TPL}}$ for a given $\beta$ by the  
linear interpolation in $(a/L)$ using the data on the 
lattices $L/a= \{8, 10\}$, $\{12,16\}$ and $\{16, 20\}$, respectively.
To estimate the systematic error of these interpolations, 
we also performed the linear interpolation in $(a/L)^2$, 
and found that the difference in the results with interpolations in $a/L$ and $(a/L)^2$ is negligible.
Furthermore, we will also show the step scaling with $s=2$ in which there is no interpolation of $L/a$, and the difference of the result should give an indirect estimation of the systematic error from the interpolation in $L/a$.

\begin{figure}[h]
\begin{center}
  \includegraphics*[height=6cm]{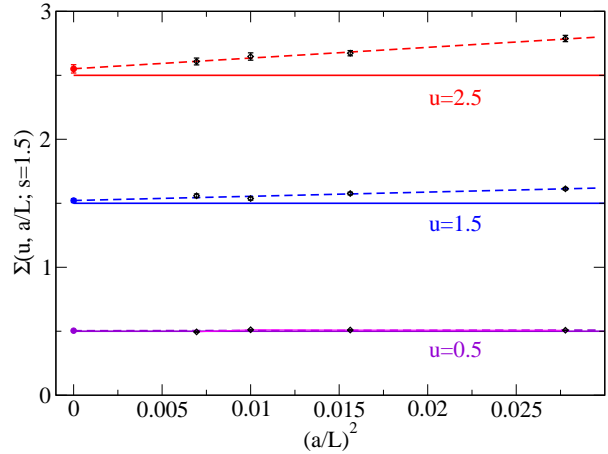} 
\end{center}
\caption{Continuum extrapolation for the case of 
input coupling $u=0.5$, $1.5$ and $2.5$ from the bottom to top respectively. 
Each solid line denotes the input renormalized coupling $u$, and the dashed line with the corresponding color denotes the linear extrapolation function of $(a/L)^2$.}
\label{fig:cont-lim-global}
\end{figure}
In Fig.~\ref{fig:cont-lim-global}, we show the examples of the 
continuum extrapolation for obtaining $\sigma(u)$ 
in the weak, intermediate and strong coupling regions.
We determine the central value of $\sigma(u)$ by the linear extrapolation in $(a/L)^2$ with four points; 
$L/a=6, 8, 10, 12$ since the leading discretization error is of $O(a^2)$ in this scheme. 
Note that, in the strong coupling region, each lattice data 
$\Sigma(u,a/L; s=1.5)$ (black data point) is quite larger than $u$, however, 
in the continuum limit, $\sigma(u)$ gets close to $u$.
This indicates that it is very important to take the continuum 
limit carefully in this kind of analysis.
We explain the reason why this continuum extrapolation is chosen as the best in the analyses in Appendix~\ref{sec:app-cont-lim}.
We also discuss the taste breaking in the continuum limit in Appendix~\ref{sec:Eigen}.


\begin{figure}[h]
\begin{center}
\includegraphics*[height=6cm]{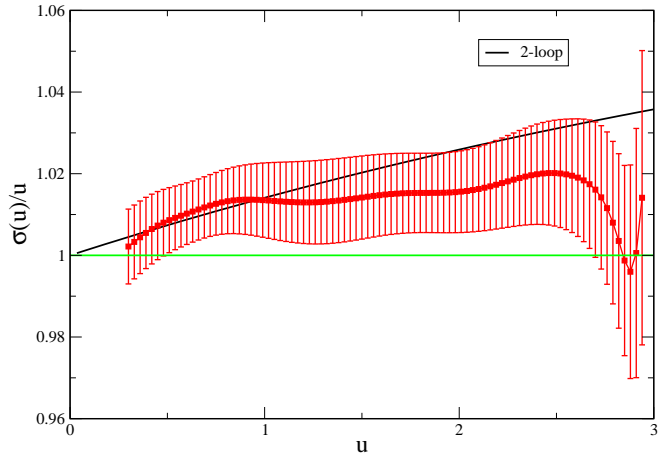}
\end{center}
\vspace{-0.5cm}
\caption{The growth rate $\sigma(u)/u$ as  a function of $u$ with statistical error. 
Two-loop perturbative value (black line) is also plotted for comparison. The horizontal (green) line denotes unity line, where the beta function is consistent with zero.} 
\label{fig:sigma-u-global}
\end{figure}
Now, we obtain the step scaling function explained above 
in a wide range of $u$. 
Figure~\ref{fig:sigma-u-global} shows the growth rate of the renormalized coupling ($\sigma(u)/u$) as a function of 
$u$ with statistical error which is estimated by jackknife method. 
We also carried out the bootstrap analysis independently, and found that the results are consistent with this jackknife analysis.

 We found two things from this plot.
The first one is that the result is consistent with perturbation theory in the weak coupling regime. 
The TPL coupling coupling with this lattice set up looks promising under this analysis method.
The other one is the central value of $\sigma(u)/u$ becomes unity around $u = 2.7$, demonstrating the signal of a fixed point.
This is the first zero of the beta function from the asymptotically free regime.
It suggests the existence of an {\it infrared} fixed point around the region. 
Unfortunately, the upper values of the error bars do not cross the line $\sigma(u)/u=1$.
This means that we cannot exclude the possibility for the coupling constant to continue growing within the error bar.
We will investigate this quantity again by focusing only the strong coupling region and adding the data.
Furthermore, will give an estimation of the systematic error of the fixed point coupling in the next subsection.

\subsection{Low energy behavior and stability of the IR fixed point} \label{sec:Nf=12-local-fit}
In the previous subsection, we found a signal of the IRFP around $u = 2.7$ from the global fit of the data.
Now we focus on the strong coupling region and will determine the fixed point coupling and the related universal quantity.
In this subsection, we take a narrow $\beta$ range in which $\beta$-dependence of $g^2_{\mathrm{TPL}}$ can be approximated by linear or quadratic functions of $\beta$.
We add more data, which is a part of the data shown in Appendix~\ref{sec:app-Taiwan-data}, to obtain the precise result and discuss the systematic uncertainties of the IRFP.

Practically, we will carry out the step scaling again with the data only in low $\beta$ region $u \ge 2.0$.
This region roughly corresponds to the range $\beta \le 7.0$.
We choose the fit range for each lattice size as shown in Table~\ref{table:local-fit-range}.
\begin{table}
\begin{center}
\begin{tabular}{|c|c|c||c|c|c|}
\hline
$L/a$ & $\beta$ range &  \# of the data &$L/a$ & $\beta$ range & \# of the data \\
\hline \hline
6 & $4.3 \le \beta \le 7.0 $  & 17 & 9 & $4.5 \le \beta \le 7.0$  & 17 \\
\hline
8 & $4.5 \le \beta \le 8.0$  & 27 & 15 & $5.3 \le \beta \le 8.0$  & 16 \\
\hline
10 & $4.5 \le \beta \le 8.0$  & 27 & 18 & $5.5 \le \beta \le 9.0$  & 7 \\
\hline
12 & $4.5 \le \beta \le 8.0$  & 21 & & &\\
\hline
\end{tabular}
\caption{The fit range of the local fit analysis for each lattice size. The data $L=9,15$ and $18$ are obtained by the $L/a$ interpolation at the fixed $\beta$ value as explained in Sec.~\ref{sec:Nf=12-global-fit}.}\label{table:local-fit-range}
\end{center}
\end{table}
The fitting function is chosen as a simple unconstraint polynomial function:
\begin{equation}
g^2_{\mathrm{TPL}}(\beta,a/L) = \textstyle{\sum_{i=0}^{N-1} \tilde{C}_i (a/L)/ \beta^{i}},\label{eq:beta-local-fit}
\end{equation}
where $N$ is the degree of the polynomial and here we take $N=3$ for $L/a=6,8$ and $N=2$ for the other lattice size.
\begin{figure}[h]
\begin{center}
  \includegraphics*[height=6cm]{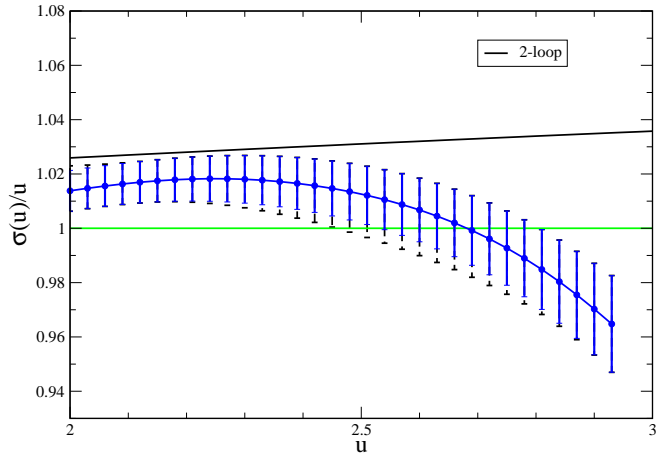} 
\caption{The local fit result for the growth rate of the TPL coupling. The solid (blue) error bar denotes the statistical error and the dot (black) error includes the systematic error. Two-loop perturbative value (black line) is also plotted for comparison. The horizontal (green) line denotes unity line, where the beta function is consistent with zero.}
\label{fig:sigma-u-local}
\end{center}
\end{figure}
We derived the step scaling function by using the same procedure as in the previous subsection.
The growth rate of the step scaling function is shown in Fig.~\ref{fig:sigma-u-local}.
As a central analysis with solid blue error bar, we take the four point linear extrapolation in $(a/L)^2$ with statistical error estimated by the jackknife method.
The dot (black) error bar includes the systematic error, which we will discuss later.
This local fit result clearly crosses the line $\sigma(u)/u=1$, which shows the existence of the IRFP.

\begin{figure}[h]
\begin{center}
  \includegraphics*[height=6cm]{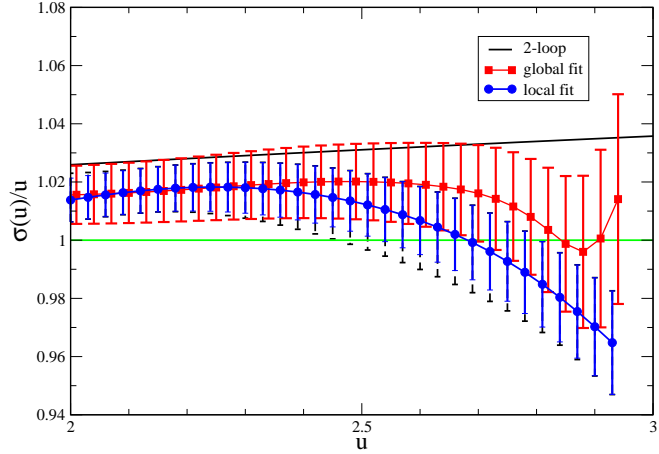} 
\caption{The comparison of the results with the global fit and those with the local fit. Two-loop perturbative value (black line) is also plotted for comparison. The horizontal (green) line denotes the line $\sigma(u)/u=1$, where the beta function is consistent with zero.}
\label{fig:sigma-u-local-global}
\end{center}
\end{figure}
Figure~\ref{fig:sigma-u-local-global} shows the comparison between the local fit result with the global fit result.
These two central value are consistent with each other within $1$-$\sigma$, despite the change of the data set, the fit range, and the fitting function.
The results strongly consolidate the existence of the stable fixed point around $u = 2.7$.
The error bar for the local fit analysis is smaller owning to the additional precise data.
We also report the data set dependence and the fit range dependence independently in Appendix~\ref{sec:several-local-fit}.

Now, we would like to estimate the systematic error in our analysis.
There are two possible dominant sources of the systematic error. 
One is from the choice of the fit range for the $\beta$-interpolation (Eq.~({\ref{eq:beta-local-fit}})).
As shown in Fig.~\ref{fig:sigma-u-local-global}, there is a small difference between the global fit and the local fit in Table~\ref{table:local-fit-range}, and we also investigate narrower range of the $\beta$.
Even if we take only the data of $\beta \le 7.0$ for all lattice size, the fitting function does not show a large difference.
We can conclude the systematic error from the choice of the fit range is small in this analysis (see Fig.~\ref{fig:several-local-sigma-u} in Appendix~\ref{sec:several-local-fit}).

\begin{figure}[h]
\begin{center}
  \includegraphics*[height=8cm]{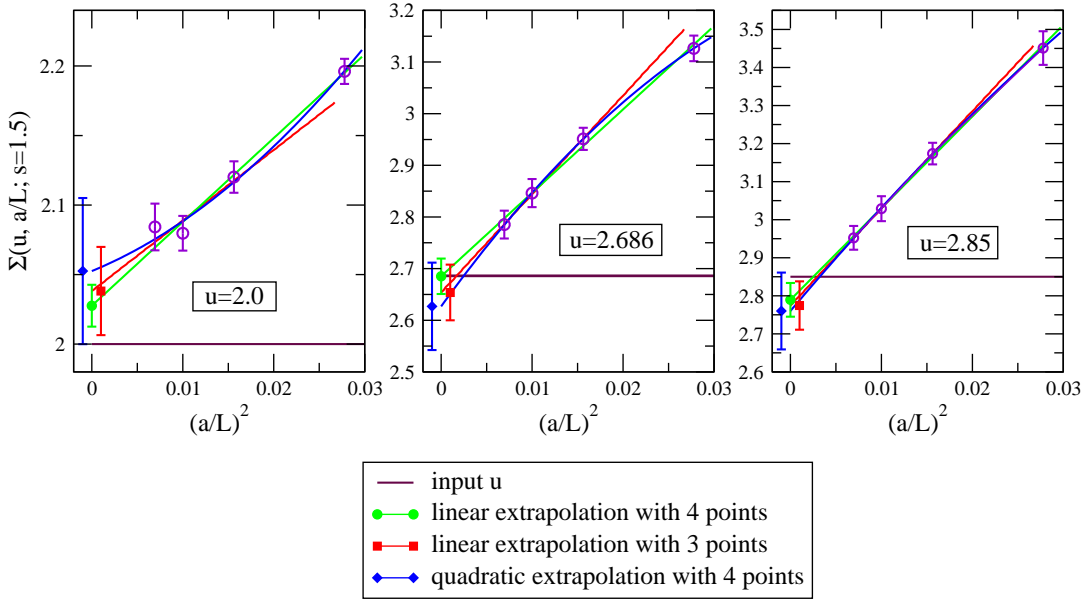} 
\end{center}
\caption{Continuum extrapolation for the case of 
input couplings $u=2.0$, $2.686$ and $2.85$. Each green line and the blue curve denotes the $4$ points linear  and quadratic extrapolation functions in $(a/L)^2$ respectively.
The red line shows the extrapolation function linear in $(a/L)^2$ for $3$ data points without the coarsest lattice data.
In the case of $u=2.0$, the step scaling function is larger than the input value, however, it becomes consistent with $u$ at $u=2.686$ and for the larger $u$ it is smaller than the input renormalized coupling constant.}
\label{fig:cont-lim-local}
\end{figure}
The other dominant systematic error comes from the continuum extrapolation.
In Fig.~\ref{fig:cont-lim-local}, we show the comparisons of several types of continuum extrapolation for $u=2.0, 2.686$ and $2.85$.
As the central value, we take the linear extrapolation in $(a/L)^2$ for $L/a=6, 8, 10, 12$. 
We estimate the systematic error by taking the difference between the central value  and the result from linear extrapolation without the data on the coarsest lattice $L/a=6$.
Furthermore we compare the central value with the quadratic extrapolation with all the data at four values of $L/a$.
Figure~\ref{fig:cont-lim-local} shows the TPL renormalized coupling has 
a small systematic error in the strong coupling region, 
and all the values in the continuum limit agree within 1-$\sigma$ statistical errors.
The total error in Fig.~\ref{fig:sigma-u-local} is estimated by adding the difference between the continuum extrapolations as a systematic error to the statistical error in quadrature.
We conclude that the existence of the IRFP is stable in this analysis.

Finally the renormalized coupling at the IRPF is
\begin{equation}
u^{\ast} =2.69 \pm 0.14\, (\mbox{stat.}) ^{+0}_{-0.16}\, (\mbox{syst.}).\label{eq:u_star}
\end{equation}
Here, the jackknife error of the running coupling is used as a 
statistical error and we estimated the systematic error coming from the continuum extrapolation.
The corresponding $\beta$ value for each lattice size at $u^{\ast}=2.69$ 
can be calculated from the $\beta$ interpolation function in 
Eq.~(\ref{eq:beta-fit}):
($\beta, L/a$)=(4.91,6),
(5.41,8), 
(5.65,10), 
(5.79,12), 
(5.91,16),
(5.94,18), 
(5.94,20).
These are the parameter sets which reproduce conformal physics after taking the continuum limit.

In addition, we mention the numerical stability of our results. 
For this purpose, we perform another step-scaling analysis based on
$s=2$ with $L/a=6, 8, 10$. 
The continuum limit is taken by linearly extrapolating three points in $(a/L)^2$.
The growth rate of the renormalized coupling is shown in Fig.~\ref{fig:several-local-sigma-u} in Appendix~\ref{sec:several-local-fit}.
We find that the running behavior in $s=2$ step scaling 
is also consistent with that in the case of $s=1.5$, and the IRFP is found 
at $u^{\ast} =2.49 \pm 0.19$ (stat.). This consolidates 
the existence of the IRFP in this theory.

\subsection{Critical exponent}\label{sec:critical-exp}
Finally we will derive the critical exponent at the IRFP.
In this theory, we have one irrelevant parameter, which is the renormalized coupling constant, around the nontrivial fixed point.
The value of the renormalized coupling at the fixed point depends on the renormalization scheme.
If we denote the scheme transformation from one renormalization scheme $u_1$ to some other one $u_2=F(u_1)$, then in the perturbative region, the function $F(u_1)$ can be expanded as a polynomial function.
The beta function for the renormalized coupling is universal up to two-loop order, however, the nonperturbatively the one for $u_2$ scheme is related with the other one as follows: 
\beq
\beta(u_2)= \frac{\partial F(u_1)}{\partial u_1} \beta(u_1). \label{eq:beta-scheme}
\eeq
In the vicinity of the IRFP, the beta function can be approximated by
\beq
\beta(u) \simeq - \gamma_g^\ast (u^\ast -u) +{\mathcal O} ((u^\ast-u)^2).
\eeq
Although the value of renormalized coupling at the IRFP is scheme dependent, we can easily find the coefficient $\gamma_g^\ast$ is the scheme independent quantity using Eq.~(\ref{eq:beta-scheme}).

Now, we compute $\gamma_g^\ast$ from the slope of $\sigma(u)/u$ against $u$, and obtain 
$s^{-\gamma_g^\ast} = 0.79 \pm 0.11(\mbox{stat.})$ in the central analysis in the Fig.~\ref{fig:sigma-u-local}.  This leads to 
\beq
\gamma_g^\ast = 0.57^{+0.35}_{-0.31} (\mbox{stat.})^{+0} _{-0.16}\, (\mbox{syst.}),
\eeq 
where the first error is statistical error using the jackknife method and the second one is the systematic error from the continuum extrapolation estimated by the comparison to the $3$ point linear continuum extrapolation.
The value of $\gamma_g^\ast$ is sensitive to the variation of the slope, 
which causes rather large statistical error.
For the $s=2$ step scaling, the critical exponent of the beta function can be derived $\gamma_g^\ast=0.31^{+0.21}_{-0.18}(\mbox{stat.})$.
This is also consistent with our main results with $s=1.5$.

Our result is consistent with 
$\gamma^{\ast {\mathrm{2-loop}}}_g \sim 0.36$ and  
$\gamma^{\ast {\mathrm{4-loop (\overline{MS})}}}_g \sim 0.28$ 
as estimated using 2-loop and 4-loop 
($\overline{{\mathrm{MS}}}$ scheme) perturbation theory~\cite{Vermaseren:1997fq,Ryttov:2010iz,Mojaza:2010cm}.
The result in the SF scheme is $\gamma^{\ast {\mathrm SF}}_g=0.13\pm 0.03$
\cite{Appelquist:2009ty}.
Our result provided a value larger than that in the SF scheme.
We can conclude that both results are almost consistent with each other since the discrepancy of $\gamma_g^\ast$ is slightly larger than 1-$\sigma$.

Another scheme independent quantity which is interesting 
to observe is the mass anomalous dimension at the IRFP. 
That is the critical exponent for the relevant operator around the IRFP.
We will report it in a forthcoming paper~\cite{Tomiya}.

\section{Summary}
We gave the explicit definition of the TPL renormalized coupling for SU($3$) gauge theory and studied the running coupling constant in the case of quenched QCD and $N_f=12$ theories.
The definition is the extension of the SU($2$) gauge theory, and we provided the perturbative calculation to define the coefficient in the case of SU($3$).

Firstly we show the TPL running coupling constant in the case of quenched QCD.
In the theory, there is confinement/deconfinement phase transition because of the finite volume effect and we study the behavior of the TPL renormalized coupling constant in the both phases.
The TPL scheme has the remarkable property that in the extremely low energy limit the coupling constant approaches to the constant ($g^{2}_{\mathrm{TPL}} \sim 32$ in the case of SU($3$) gauge theory), when the theory is in the confinement phase.
From this analysis, the TPL coupling is found to be useful only in the deconfinement phase, so that we should study the phase structure in the parameter spaces and search for the available region before the running coupling study.
The running coupling constant is consistent with the perturbative result in high energy region, and it runs more slowly than that in the $1$-loop perturbation in the low energy region.
The running coupling constant in nonperturbative region is scheme dependent and is different from SF and Wilson loop schemes.

In the case of SU($3$) gauge theory with $12$ massless Dirac fermions in the fundamental representation, 
we inverstigated the phase structure on $\beta$--$ma$ space and the vacuum structure related with $Z_3$ center symmetry to identify the true vacua.
We revealed the phase structure for $N_f=12$ massive and massless fermion theories and found that there is a bulk phase transition near the massless region at a point of $\beta < 4.0$ region.
In such phase, the TPL coupling is not available since the theory shows the confinement behavior.
We used the configuration only in the deconfinement phase to investigate the running coupling constant, and also found that the chiral symmetry seems to be preserved.

We also discussed the vacuum structure and the center symmetry breaking in our simulation setup using the semi-classical analysis, and generated the configurations in the true vacua.
The vacuum structure depends on the boundary condition of the fermions, in the case of our definition, the configurations whose Polyakov loop in the untwisted direction has the nontrivial phase shows the minimum of the potential.

Finally, we have found a solid evidence for the existence of an IRFP using the TPL scheme. 
The coupling constant at the IRFP is $g^{*2}_{\mathrm{TPL}} \sim 2.69$.
The stability of the fixed point is discussed, and we can conclude there is the IRFP after the systematic uncertainties are included.

\section*{Acknowledgements and Comments}
This basic strategy of the paper was shown in the letter paper~\cite{Aoyama:2011ry}, which is already withdrawn from the arXiv.
Before the letter paper \cite{Aoyama:2011ry}  would be published, the collaboration was reset when a part of members released the paper~\cite{Ogawa-paper} independently, since there was scientific conflicts concerning with the  analysis method and the quality of the full paper.
This paper includes the further updated data after the collaboration was reset, and the differences of the analysis method are discussed in Appendix~\ref{sec:app-Taiwan-global} and \ref{sec:app-cont-lim}.

Some readers could not understand the reason why there are two kinds of data sets in Appendix~\ref{sec:app-Japan-data} and \ref{sec:app-Taiwan-data}.
Actually, we have not obtained the raw data of the Polyakov loop and configurations in the Appendix~\ref{sec:app-Taiwan-data} since some members have not send them for more than five months after the collaboration reset while they agreed with sharing the data. 
Consequently, the analysis of  Fig.\ref{fig:P4-L-dependence} and Appendix~\ref{sec:Eigen} have been done with only the data in Appendix~\ref{sec:app-Japan-data}.
Furthermore, as we explained the data sets in Appendix~\ref{sec:app-Taiwan-data} is strongly prejudiced for a specific region, and that is not suitable to study the global fit analysis.

We would like to thank all ex-collaborators in the proceedings \cite{Bilgici:2009nm,Itou:2010we} and the letter paper\cite{Aoyama:2011ry} for the discussions.
In particular, T.~Onogi and T.~Yamazaki gave an original idea for the TPL scheme and several important suggestions and comments.
The PHB code for the quenched QCD in the Sec.~\ref{sec:quenched-running} was developed by T.~Yamazaki.
The HMC codes were developed by H.~Matsufuru and E.~Shintani for several supercomputers, and T.~Aoyama and K.~Ogawa also gave an effort to developing the GPU code. 
A part of configuration generations had been done by M.~Kurachi, H.~Matsufuru, K.~Ogawa, H.~Ohki, T.~Onogi, 
E.~Shintani and T.~Yamazaki.
We would like to thank these people for the collaboration and would like to address that the original members of this project were T.~Onogi, M.~Kurachi and E.~I. and the collaboration started when we were in YITP, Kyoto.

We also appreciate G.~Fleming, P.~de~Forcrand, H.~Fukaya, A.~Hasenfratz, S.~Hashimoto, J.~Kuti, M.~L\"{u}scher, A.~Patella, F.~Sannino, Y.~Taniguchi, A.~Tomiya and N.~Yamada for useful discussions and comments.
And we would like to thank A.~Irie for making Fig.~\ref{fig:phase-Nf12}.
We really appreciate T.~Onogi and K.~Higashijima for encouraging to release this paper.

Numerical simulation was carried out on
NEC SX-8 and Hitachi SR16000 at YITP, Kyoto University,
NEC SX-8R at RCNP, Osaka University,
and Hitachi SR11000, SR16000 and IBM System Blue Gene Solution at KEK 
under its Large-Scale Simulation Program
(No.~09/10-22, 10-16, (T)11-12, 12-16 and 12/13-16), as well as on the GPU cluster at Osaka University and 
Taiwanese National Centre for High-performance Computing.
We acknowledge Japan Lattice Data Grid for data
transfer and storage.
E.I. is supported in part by
Strategic Programs for Innovative Research (SPIRE) Field 5.
This work is supported in part by the Grant-in-Aid of the Ministry of Education (No. 
22740173). 

\newpage
\appendix
\section{Raw data of TPL coupling constant}\label{sec:app-Japan-data}
\begin{table}[h]
\begin{center}
\begin{tabular}{|r||r|r||r|r||r|r||}
\hline
& \multicolumn{2}{|c||}{$L/a=6$}
& \multicolumn{2}{|c||}{$L/a=8$}
& \multicolumn{2}{|c||}{$L/a=10$}\\
\hline
\multicolumn{1}{|c||}{$\beta$}
& \multicolumn{1}{c|}{$g^2_{\mathrm{TPL}}$}  &  \# of Trj. 
& \multicolumn{1}{c|}{$g^2_{\mathrm{TPL}}$}  &  \# of Trj.
& \multicolumn{1}{c|}{$g^2_{\mathrm{TPL}}$}  &  \# of Trj. \\
\hline
100.0& 0.06304(31) &  49500 &             &         &             &          \\
99.0 &             &        & 0.06369(27) &   59500 & 0.06389(38) &   39500  \\          
50.0 & 0.13229(49) &  44000 & 0.13084(83) &   49500 & 0.13349(93) &   65500  \\               
20.0 &  0.3895(22) &  72000 &  0.3910(31) &   73000 &  0.3824(54) &   49000  \\               
18.0 &  0.4512(34) &  60000 &  0.4565(39) &   83500 &  0.4513(60) &   74000  \\               
16.0 &  0.5158(31) & 108000 &  0.5282(53) &   78500 &  0.5345(66) &   88000  \\             
15.0 &  0.5739(44) &  80000 &  0.5849(53) &  100000 &  0.5863(68) &   98500  \\            
14.0 &  0.6274(48) &  90000 &  0.6492(66) &   95500 &  0.6359(78) &  120000  \\               
13.0 &  0.6944(59) &  80000 &  0.7156(63) &  106000 &  0.7357(93) &  102500  \\              
12.0 &  0.7844(67) &  90000 &  0.7930(75) &  126500 &  0.8205(15) &   80000  \\               
11.0 &  0.9154(93) &  80000 &  0.9210(98) &  126000 &  0.9939(19) &  102500  \\               
10.0 &   1.050(14) &  54000 &   1.071(18) &   72000 &   1.129(25) &   83500  \\            
9.5  &   1.120(13) &  80000 &   1.161(15) &  100000 &   1.172(23) &   79600  \\         
9.0  &   1.225(15) &  78000 &   1.279(15) &  130500 &   1.352(29) &   80000  \\         
8.5  &   1.303(15) &  80000 &   1.380(18) &  104400 &   1.433(28) &  101000  \\         
8.0  &   1.530(19) &  78000 &   1.570(35) &   63500 &   1.612(40) &   68000  \\        
7.5  &   1.603(19) &  80000 &   1.710(27) &   93600 &   1.770(33) &   91000  \\        
7.0  &   1.812(23) &  54000 &   1.924(22) &  153000 &   1.987(31) &  208000  \\  
6.7  &   1.893(23) &  80000 &   2.079(30) &  112000 &   2.078(39) &  108200  \\         
6.5  &   1.968(32) &  54000 &   2.109(30) &   99000 &   2.226(43) &  140000  \\        
6.3  &   2.108(29) &  80000 &   2.235(27) &  104000 &   2.336(45) &   99400  \\       
6.0  &   2.206(25) &  90000 &   2.383(42) &   95000 &   2.476(40) &  130000  \\       
5.7  &   2.339(29) &  80000 &   2.503(34) &  110000 &   2.589(51) &   80000  \\        
5.5  &   2.436(31) &  72000 &   2.660(50) &   75000 &   2.804(56) &  114800  \\        
5.3  &   2.546(29) &  80000 &   2.765(39) &  113000 &   2.908(57) &   80000  \\      
5.0  &   2.716(37) &  96000 &   2.917(61) &   94000 &   3.147(69) &   95000  \\      
4.7  &   2.790(39) &  78000 &   3.101(52) &   85000 &   3.380(72) &   99200  \\          
4.5  &   2.810(35) & 108000 &   3.219(55) &  113000 &   3.599(76) &  220400  \\
4.3  &   2.942(34) &  94000 &             &         &             &          \\
4.0  &   2.888(48) &  69000 &             &         &             &          \\
\hline
\end{tabular}
\caption{$L/a=6$, $8$, and $10$.(Data set A)} \label{tab:g2L6-8-10}
\end{center}
\end{table}
\newpage

\begin{table}[h]
\begin{center}
\begin{tabular}{|r||r|r||r|r||r|r||}
\hline
& \multicolumn{2}{|c||}{$L/a=12$}
& \multicolumn{2}{|c||}{$L/a=16$}
& \multicolumn{2}{|c||}{$L/a=20$}\\
\hline
\multicolumn{1}{|c||}{$\beta$}
& \multicolumn{1}{c|}{$g^2_{\mathrm{TPL}}$}  &  \# of Trj. 
& \multicolumn{1}{c|}{$g^2_{\mathrm{TPL}}$}  &  \# of Trj.
& \multicolumn{1}{c|}{$g^2_{\mathrm{TPL}}$}  &  \# of Trj. \\
\hline
99.0 & 0.06381(48) &   36000 & 0.06331(72) &    27900 & 	   &          \\
50.0 &  0.1316(11) &   64800 &  0.1327(15) &    60900 & 0.1336(13) &   147100 \\        
20.0 &  0.3927(52) &   86400 &  0.3956(62) &    85800 & 0.4002(79) &   123900 \\        
18.0 &  0.4463(61) &   90000 &   0.469(14) &    40000 & 0.4509(83) &   148400 \\        
16.0 &  0.5463(97) &   84600 &  0.5434(96) &   116000 &  0.547(11) &   155500 \\      
15.0 &  0.6110(12) &   57200 &   0.601(12) &   100600 &            &          \\
14.0 &  0.6478(13) &   79200 &   0.641(12) &   126000 &  0.637(13) &   125900 \\
13.0 &  0.7278(12) &   99800 &   0.746(14) &   101000 &            &          \\
12.0 &  0.8444(14) &  102000 &   0.838(16) &   118000 &  0.832(18) &   258200 \\        
11.0 &  0.9601(19) &  104000 &   0.956(23) &   109800 &            &          \\
10.0 &   1.133(22) &  159000 &   1.132(21) &   186900 &  1.173(26) &   263700 \\
9.5  &   1.229(24) &  128200 &   1.275(21) &   227400 &            &          \\
9.0  &   1.315(24) &  160500 &   1.376(33) &   219100 &  1.427(33) &   322400 \\         
8.5  &   1.479(27) &  160600 &   1.523(27) &   321000 &            &          \\
8.0  &   1.589(33) &  104400 &   1.633(29) &   379500 &  1.696(38) &   295300 \\
7.5  &   1.813(36) &  169200 &   1.881(38) &   279900 &            &          \\
7.0  &   2.058(40) &  162400 &   2.112(45) &   287700 &  2.077(45) &   430700 \\      
6.7  &   2.168(35) &  162600 &             &          &            &          \\
6.5  &   2.298(40) &  212000 &   2.276(46) &   466900 &  2.370(56) &   301400 \\    
6.3  &   2.373(42) &  181200 &             &          &            &          \\
6.0  &   2.498(46) &  180000 &   2.662(57) &   213600 &  2.625(67) &   443700 \\
5.7  &   2.678(46) &  162800 &   2.860(57) &   293400 &  2.901(60) &  1892800 \\
5.5  &   2.824(93) &   64400 &   3.044(59) &   387000 &   3.05(28) &   262800 \\
5.3  &   2.974(52) &  191700 &   3.055(70) &   457000 &            &          \\
5.0  &   3.235(64) &  241800 &             &          &            &          \\
4.7  &   3.600(70) &  262200 &             &          &            &          \\
4.5  &    3.57(12) &  269500 &             &          &            &          \\
\hline
\end{tabular}
\caption{$L/a=12$, $16$, and $20$.(Data set A) (The data for $\beta=5.5, L/a=12$ becomes poor statistics since the bugged data was found after the simulations.)} \label{tab:g2L12-16-20}
\end{center}
\end{table}

\newpage
\section{Additional data set of TPL coupling constant for the local fit analysis}\label{sec:app-Taiwan-data}
\begin{table}[h]
\begin{center}
\begin{tabular}{|r||r|r||r|r||r|r||}
\hline
& \multicolumn{2}{|c||}{$L/a=6$}
& \multicolumn{2}{|c||}{$L/a=8$}
& \multicolumn{2}{|c||}{$L/a=10$}\\
\hline
\multicolumn{1}{|c||}{$\beta$}
& \multicolumn{1}{c|}{$g^2_{\mathrm{TPL}}$}  &  \# of Trj. 
& \multicolumn{1}{c|}{$g^2_{\mathrm{TPL}}$}  &  \# of Trj.
& \multicolumn{1}{c|}{$g^2_{\mathrm{TPL}}$}  &  \# of Trj. \\
\hline
20.13 &  0.3892(17)  &  330200   &   0.3907(26) &   188800 &  0.3975(42) &    85500 \\
17.55 &  0.4645(18)  &  353800   &   0.4658(31) &   166700 &  0.4740(62) &    88400 \\
15.23 &  0.5647(29)  &  339500   &   0.5733(49) &   170400 &  0.5671(86) &    91600 \\
13.85 &  0.6420(35)  &  352300   &   0.6522(55) &   190000 &  0.6563(93) &    83800 \\
11.15 &  0.8818(55)  &  330100   &   0.8971(98) &   164100 &   0.940(18) &   112600 \\
9.42  &  1.1443(60)  &  389300   &    1.191(10) &   317200 &   1.237(14) &   274700 \\
8.45  &  1.3488(84)  &  289100   &   1.4204(70) &   987600 &   1.444(14) &   503900 \\
7.82  &  1.5313(91)  &  383800   &    1.601(13) &   344500 &   1.650(14) &   454300 \\
7.80  &              &           &              &          &   1.717(56) &    45300 \\
7.25  &              &           &              &          &   1.890(58) &    54100 \\
7.11  &              &           &    1.842(11) &   720500 &   1.956(25) &   256700 \\
7.10  &              &           &    1.866(32) &    78200 &             &          \\
6.85  &              &           &              &          &   2.068(60) &    53900 \\
6.80  &              &           &    1.984(34) &    78000 &             &          \\
6.76  &   1.869(12)  &  306400   &    2.003(18) &   356600 &   2.086(25) &   305900 \\
6.55  &              &           &    2.084(37) &    78000 &             &          \\
6.47  &              &           &   2.1360(10) &  1293000 &   2.220(25) &   307200 \\
6.45  &              &           &              &          &   2.380(69) &    53700 \\
6.25  &              &           &    2.313(47) &    78100 &             &          \\
6.15  &              &           &              &          &   2.441(66) &    47700 \\
6.12  &  2.1429(95)  &  603000   &    2.307(18) &   584700 &   2.434(35) &   249700 \\
5.95  &              &           &    2.317(64) &    74800 &             &          \\
5.90  &              &           &              &          &   2.528(77) &    47300 \\
5.81  &   2.248(11)  &  530200   &    2.471(22) &   415800 &   2.605(37) &   216800 \\
5.80  &              &           &    2.516(47) &    78200 &             &          \\
5.53  &   2.408(11)  &  718600   &    2.676(29) &   471800 &   2.784(46) &   177000 \\
5.36  &   2.489(10)  &  696700   &    2.698(66) &    42900 &   2.820(65) &    89400 \\
\hline
\end{tabular}
\caption{$L/a=6$, $8$, and $10$.(Data set B)} \label{tab:g2L6-8-10-Taiwan}
\end{center}
\end{table}
\newpage

\begin{table}[h]
\begin{center}
\begin{tabular}{|r||r|r||r|r||}
\hline
& \multicolumn{2}{|c||}{$L/a=12$}
& \multicolumn{2}{|c||}{$L/a=16$}\\
\hline
\multicolumn{1}{|c||}{$\beta$}
& \multicolumn{1}{c|}{$g^2_{\mathrm{TPL}}$}  &  \# of Trj. 
& \multicolumn{1}{c|}{$g^2_{\mathrm{TPL}}$}  &  \# of Trj. \\
\hline
20.13  &   0.3982(63) &    83000 & 0.4138(75) &    79800 \\
17.55  &   0.4662(68) &    85100 & 0.4780(92) &    83600 \\
15.23  &    0.588(12) &    75200 &  0.566(14) &    80700 \\
13.85  &    0.674(14) &    79100 &  0.699(19) &    70800 \\
11.15  &    0.914(13) &   173700 &  0.962(30) &    72500 \\
9.42   &    1.263(19) &   256200 &  1.228(29) &   147600 \\
8.45   &    1.470(18) &   397200 &  1.520(25) &   368200 \\
7.82   &    1.670(26) &   262300 &  1.695(52) &   136300 \\
7.11   &    1.966(30) &   256800 &  1.996(40) &   244700 \\
6.76   &    2.095(34) &   265700 &  2.163(65) &   136400 \\
6.47   &    2.278(29) &   330900 &  2.391(50) &   273100 \\
6.12   &    2.562(36) &   283200 &  2.489(62) &   183700 \\
5.81   &    2.723(45) &   269100 &  2.729(79) &   186000 \\
5.53   &    2.950(56) &   233600 &  2.953(83) &   191200 \\
5.36   &    3.030(50) &   272600 &   3.06(10) &   187200 \\
\hline
\end{tabular}
\caption{$L/a=12$ and $16$. (Data set B)} \label{tab:g2L12-16-20-Taiwan}
\end{center}
\end{table}

\section{Data set dependence, fit range dependence and the step scaling size dependence}\label{sec:several-local-fit}
We show the two kinds of result for the growth rate from the global fit analysis and local fit analysis in Sec.~\ref{sec:Nf=12-global-fit} and \ref{sec:Nf=12-local-fit} respectively.
They are consistent with each other within $1$-$\sigma$ although they have the difference data sets, the different fit range and fitting function each other.
Here we would like to show the each systematic uncertainties.

\begin{figure}[h]
\begin{center}
\includegraphics*[height=10cm]{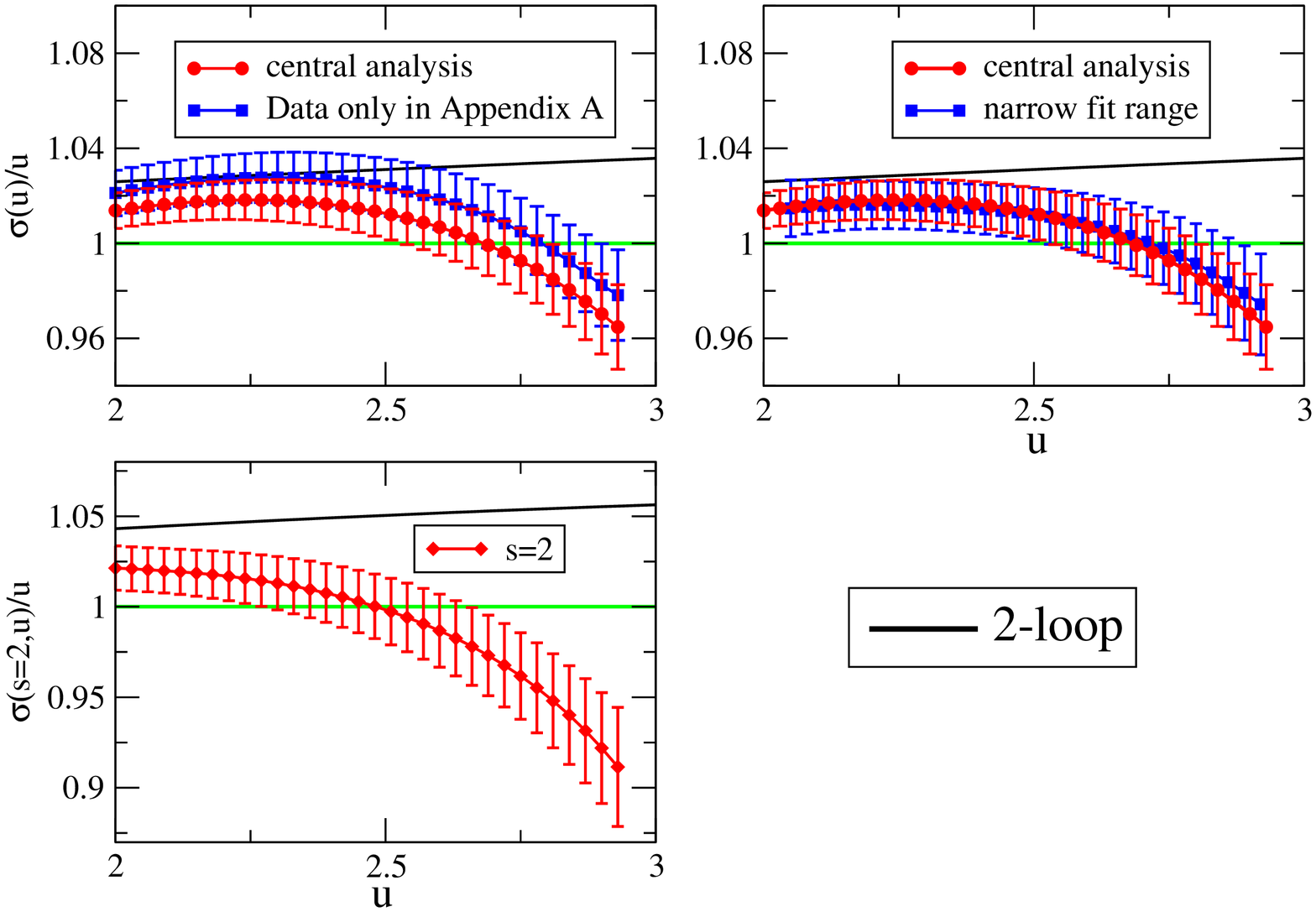}
\caption{The comparison for several local fit result. The top left panel shows the data set dependence. The top right panel shows the fit range dependence. The bottom panel shows the $s=2$ step scaling, and then the comparison for the position of $u^\ast$ shows the step scaling parameter dependence.}
\label{fig:several-local-sigma-u}
\end{center}
\end{figure}
The left top panel in Fig.~\ref{fig:several-local-sigma-u} shows the data set dependence.
The central analysis includes both Data sets in Appendix~\ref{sec:app-Japan-data} and \ref{sec:app-Taiwan-data}, while the blue result is obtained by only Data sets in Appendix~\ref{sec:app-Japan-data}, which is the same data sets with the global fit analysis.
The results are consistent with each other within $1$-$\sigma$.
The right top panel in the Fig.~\ref{fig:several-local-sigma-u} shows the fit range dependence.
The blue result is obtained by the data in the narrow $\beta$ range; $\beta \le 7.0$ for all lattice size.
We can find the result is completely consistent with each other, and this local fit analysis is quite stable under the change of the fit range.
Actually, we focus on the local $\beta$ region where all data can be fitted by the linear or quadratic function in $\beta$.
Such results can be expected when the data can be fitted well.

The bottom panel in Fig.~\ref{fig:several-local-sigma-u} shows the result of $s=2$ step scaling.
Although the step scaling function depends on the step scaling parameter, the comparison for the position $u^\ast$ must be independent of the step scaling parameter.
We can find the position is consistent with that for $s=1.5$, so that we can confidently conclude the existence of the fixed point and the interpolation works well in the $s=1.5$ step scaling.

\section{The effect of the nonuniform data on the global fit}\label{sec:app-Taiwan-global}
The fixed point in this paper shows $u^\ast =2.69$, however, the paper~\cite{Ogawa-paper} shows $u^\ast \sim 2.0$.
We consider that this difference comes from the mismatching the analysis method and the data sets quality in the paper~\cite{Ogawa-paper}.
In this appendix, we would like to consider the problem.

The data in the paper~\cite{Ogawa-paper} are strongly concentrated around $\beta = 5.0$ and they are a part of Data set A and all data of set B.
In the paper~\cite{Ogawa-paper}, the authors carried out the global fit analysis.
As we mentioned in our guiding principle point $1$, when we use the global fit in the broad $\beta$ region by using a single fitting function, ideally the data do not favor a specific region.
Our analysis for the global fit used only the data A in which each data has a similar accuracy and the interval of $\beta$ is chosen to realize the almost constant growth rate of the renormalized coupling on the lattice.
On the other hand, the data set B is strongly concentrated around $\beta= 5.0$ -- $6.0$ and a part of the data has quite high statistics in the high $\beta$ region and for the small $L/a$.
In this appendix, we would like to derive the global step scaling function by using the data in the both sets A and B and discuss the effect of the prejudiced data.

\begin{figure}[h]
\begin{center}
\includegraphics*[height=10cm]{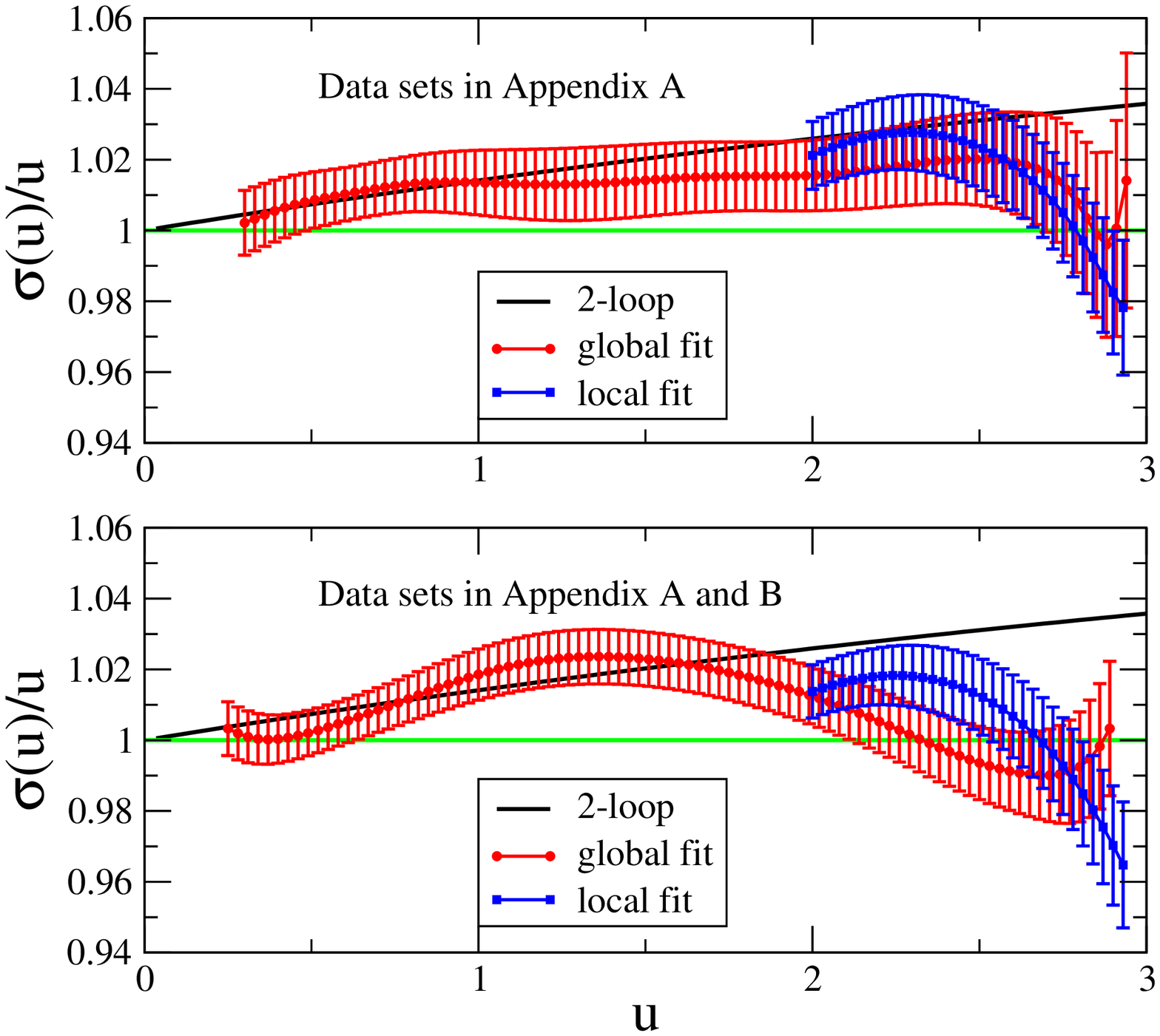} 
\caption{The data set dependence and the fit range dependence. The top panel is obtained by only the data set A, and the bottom one is by the data sets A and B. Each red data denotes the global fit analysis and the blue one shows the local fit analysis. We can find the bottom panel shows the large fit range dependence.}
\label{fig:taiwan-global}
\end{center}
\end{figure}
In Fig.~\ref{fig:taiwan-global}, the results by using the data set A and the ones by using both A and B are shown in the top and bottom panels respectively.
Each procedure of the step scaling is the same with the Sec.~\ref{sec:Nf=12-global-fit} and Sec.~\ref{sec:Nf=12-local-fit} for the global and local fit analysis respectively.
Each red result in fig.~\ref{fig:taiwan-global} denotes the global fit analysis and the blue one denotes the local fit analysis.
The latter global fit result looks very similar with the result of the paper~\cite{Ogawa-paper}.
It shows the worse matching with the perturbative result, although still it is consistent with the perturbative prediction within the statistical error.
The values $u^\ast$ becomes smaller than the other ones with more than $2$-$\sigma$ discrepancy.

Now, we should consider which result is reliable.
The global fit with the broad regime is always dangerous since the interpolated value includes non-vanishing contributions coming from the far region.
To remove such effect we also carried out the local fit only with the focusing regime.
The comparison the global fit result and the local fit result with both data A and B shows the larger fit range dependence rather than our central analysis in Fig.~\ref{fig:sigma-u-local-global}.
The guiding principle point $1$ must be important for such global fit analysis.
That is the reason why we did not use the whole data of the data set B, when we would like to show the global behavior of the TPL running coupling constant.

\section{Comments on the estimation method of the discretization effect}\label{sec:app-cont-lim}
The discretization effect of the step scaling function $\Sigma(u,a/L;s)$ has two origins.
The first one is a simple discretization effect of the renormalized coupling in the larger lattice size $(sL/a)$.
The second one comes from the tuned value of $\beta$ to reproduce the input quantity $u$ in the smaller lattice $(L/a)$.
When we take the continuum limit, we fixed the physical box size $L$ and the lattice spacing $a$ ($=\beta$), and then the leading term of the former ($O((a/sL)^2)$) is smaller than the later one ($O((a/L)^2)$).
So, it must be safe to avoid the interpolation of the small lattice sets if we consider the discretization effect seriously.

In our simulation, we have $L/a=6,8,10,12,16$ and $20$.
Then in the $s=1.5$ step scaling we can take $4$ data points to estimate the $O(a^2)$ effect with avoiding the small $L/a$ interpolation, on the other hand $s=2$ step scaling has only $3$ data points.
One of the advantages to use $s=1.5$ step scaling is that there is the finest lattice data($L/a=12$).
Furthermore the chi square fit with only one degree of freedom is strongly disturbed the statistical fluctuation, then we take the $4$ points linear extrapolation as the central analysis in this paper.

In the paper~\cite{Ogawa-paper}, they carried out the interpolation for $L/a=7$ by using $L/a=6$ and $8$.
However, the interpolation is dangerous since it is the coarsest two lattices interpolation.
Actually, the raw data of the TPL (Figs.~\ref{fig:global-g2}) shows the largest difference between $L/a=6$ and $8$ in the low $\beta$ region and that must induce a large uncertainty of the interpolation.
Furthermore, the estimation of the systematic uncertainty between $4$ data points linear extrapolation for $L/a=6,7,8,10 \rightarrow 12,14,16,20$ and $3$ data points linear extrapolation for $L/a=7,8,10\rightarrow 14,16,20$ is nonsense, since the data of $L/a=7$ is generated by $L/a=6$ lattice data and thus the later extrapolation does not remove the effects of the coarsest lattice.

\section{Eigenvalue of the Dirac operator}\label{sec:Eigen}
In this appendix, we report results for the eigenvalue distribution of the Dirac operator.
We confirm two things from the quantity.
At first, we show the global shape of the eigenvalue distribution.
We find that the data in the weak coupling region, where the perturbative theory gives good approximations, is consistent with the tree level analysis in Sec.~\ref{sec:pert-eigen}.
Furthermore, the $\beta$ dependence of the data is smooth in the whole region and the lowest eigenvalue is nonzero even in the lowest $\beta$ in our simulation parameter.
These observations also indicate that the theory is in the deconfinement and chiral symmetric phase.
Secondly, we discuss the taste breaking in our simulation in Sec.~\ref{sec:taste-breaking}.
In the case of the high $\beta$ and the large lattice extent, the raw data of the low lying eigenvalues shows the degeneracy of the taste.
In the strong coupling region, we find that the effect of the taste breaking becomes mild in the continuum limit.
We consider up to the order $a^4$ for the continuum extrapolation, which is the same order as for the running coupling study.
How much the taste breaking recovers up to this order would be an indirect estimation for the discretization effect for the TPL coupling constant.

This section is a preliminary report for the study on the eigenvalues of the Dirac operator.
We will report the detailed studies in the independent paper in near future\cite{Tomiya}.

\subsection{Perturbative analysis}\label{sec:pert-eigen}
Let us consider the  massless staggered-Dirac operator $D(x,y)$,
\beq
D(x,y)= \sum_\mu \eta_\mu (x) \left[ U_\mu (x) \delta_{x+\hat{\mu},y} -U^\dag_\mu(x-\hat{\mu}) \delta_{x-\hat{\mu},y} \right],
\eeq
where $\eta_\mu(x)$ is the staggered phase.
This eigenvalues of $D$ are pure imaginary since it is the anti-hermitian $D^\dag(x,y)=-D(x,y)$:
\beq
 D(x,y) \psi^{(l)}_\lambda=i \lambda^{(l)} \psi^{(l)}_\lambda,
\eeq
where $\psi^{(l)}_\lambda$ denotes a Dirac fermion and $(l)$ denotes the level of the eigenvalues.
We define the lowest one as $l=1$.
The degree of freedom of one staggered-Dirac operator is $16$, and there are additional $3$ color and $3$ smell indices for our Dirac fermion (see Eq.~(\ref{eq:fermion-bc})).
The number of flavor ($N_f=12$) is realized by ($4$ taste's) $\times$ ($3$ smell's) degrees of freedom and there are $3$ flavor ($=$smell) staggered Dirac fermions.
The operator $D^\dag D$ is hermitian and positive definite, and it can be decomposed to the operators on the even and odd sites.
In this work, we measure the eigenvalues only positive and in the even-to-even sites, and then the degeneracy of one staggered-Dirac operator ($4$ taste's$\times$ $4$ spinor's) becomes half.
The remain degrees of freedom ($3$ color's $\times$ $3$ smell's) can be counted as the unphysical twisted momenta in the twisted directions as explained in the Sec.~\ref{sec:vacuum}.

\begin{figure}[h]
\begin{center}
\includegraphics*[height=6cm]{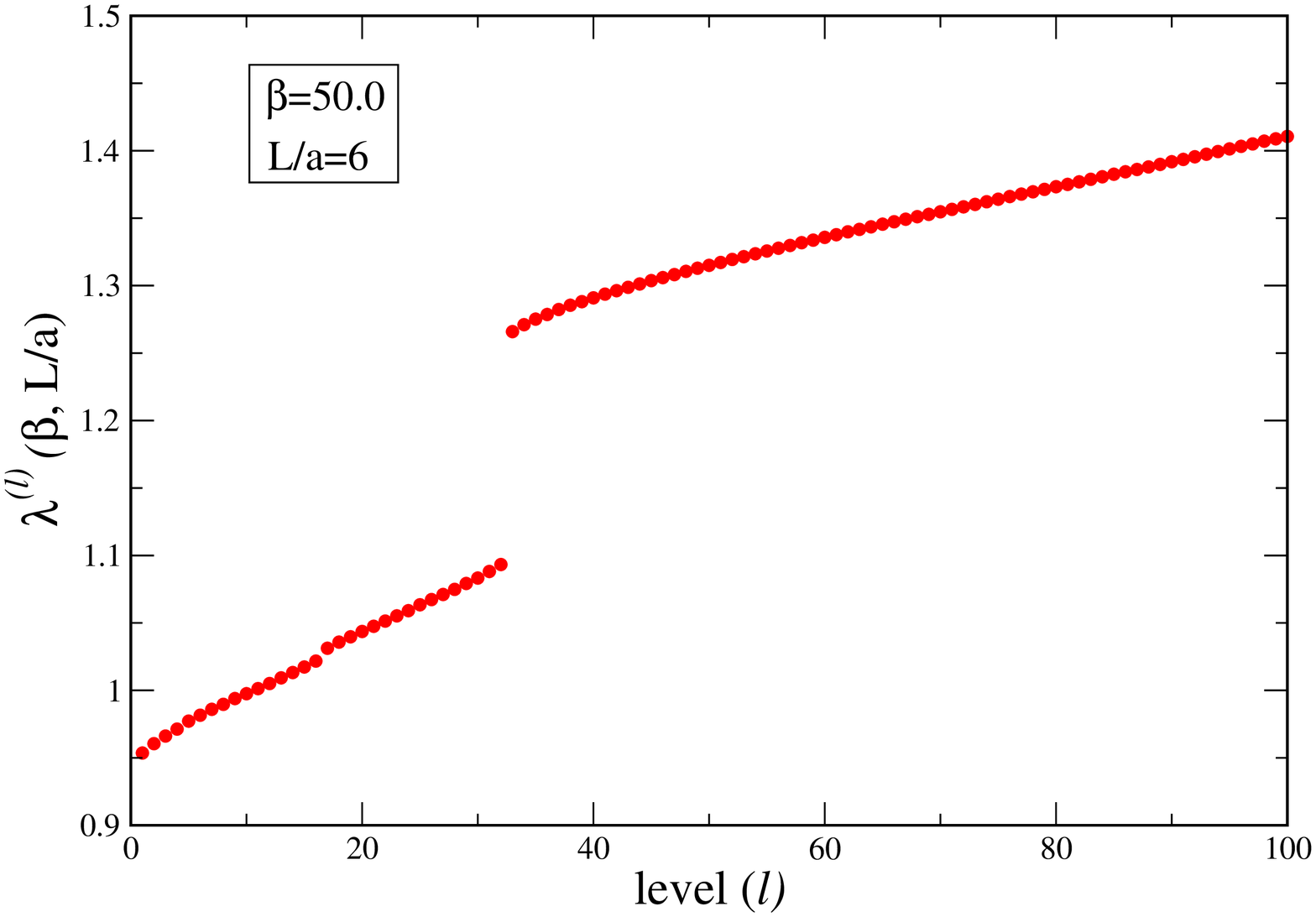} 
\caption{The eigenvalues for $\beta=50,L/a=6$. The statistical error is the same size of the symbol.}
\label{fig:eigen-beta-50-L-6}
\end{center}
\end{figure}
At the tree level, we can calculate the eigenvalue of the Dirac operator on the lattice:
\beq
\lambda^2 =  4 \sum_\mu \sin^2 \frac{\hat{k}_\mu}{2}, 
\eeq
where $\hat{k}_\mu$ denotes the momentum of the fermion field for each direction.
The leading order of $O(a)$ for low lying eigenvalues is proportional to the sum of $\hat{k}_\mu$.
In our simulation there is twisted boundary condition and the non-trivial vacuum phase, and then the momentum is given by the Eq.~(\ref{eq:k-mu-theta}) as $\hat{k}^{\theta}_\mu$.
The lowest momentum is given in the following case of 
\beq
(n_\mu^{ph},n_\mu^\perp)&=&(-1,2) \mbox{ or } (0,0) \mbox{  for both $\mu=x$ and $y$}, \nonumber\\
n_\mu^{ph}&=&0 \mbox{  for both $\mu=z$ and $t$}\nonumber.
\eeq
The degree of degeneracy for these momentum combinations is $4$, and the sum of $\hat{k}_\mu$ is given by $\sqrt{10} \pi/\hat{L}$.
The second lowest mode is given in the case of 
\beq
(n_\mu^{ph},n_\mu^\perp)&=&(-1,2) \mbox{ or } (0,0) \mbox{  for $\mu=x$ or $y$}, \nonumber\\
(n_\mu^{ph},n_\mu^\perp)&=&(-1,1) \mbox{ or } (0,1) \mbox{  for $\mu=y$ or $x$ (in same order)},\nonumber\\
n_\mu^{ph}&=&0 \mbox{  for both $\mu=z$ and $t$}\nonumber
\eeq
The degree of degeneracy of that is $8$, and the sum of the momentum is $\sqrt{18} \pi/\hat{L}$.
The third lowest mode is also counted, it is given by
\beq
(n_\mu^{ph},n_\mu^\perp)&=&(-1,2) \mbox{ or } (0,0) \mbox{  for both $\mu=x$ and $\mu=y$}, \nonumber\\
n_\mu^{ph}&=&0 \mbox{  for $\mu=z$ or $t$},\nonumber\\
n_\mu^{ph}&=&-1 \mbox{  for  $\mu=t$ or $z$ (in same order)}.\nonumber
\eeq
The number of degrees of degeneracy of that is $8$, and the sum of the momentum is $\sqrt{22} \pi/\hat{L}$.

Let us compare the measured value of the simulation with this tree level analysis.
The total degree of the degeneracy should be multiplied by $8$, since we measure the eigenvalue of the staggered fermion on even-to-even site as we explained.

Figure~\ref{fig:eigen-beta-50-L-6} shows the first $100$ eigenvalues in the $\beta=50$, $L/a=6$.
There is a clear gap between $l=32$ and $l=33$ as we expected, although there is a large taste breaking since the lattice size is small.
The ratio of the eigenvalue between first and $33$rd levels is $1.328(2)$, and it is almost consistent with the tree level prediction $\sqrt{\frac{18}{10}}$.
On the other hand, we cannot see the second gap, which we expect to lie between $96$th and $97$th levels.
The numerical value of the ratio of them is $1.110(1)$, and the value is completely consistent with $\sqrt{\frac{22}{18}}$.
Although the second gap is not clear, the eigenvalue reproduces the tree level prediction.

\subsection{The eigenvalue distribution and the taste breaking}\label{sec:taste-breaking}
\begin{figure}[h]
\begin{center}
\includegraphics*[height=8cm]{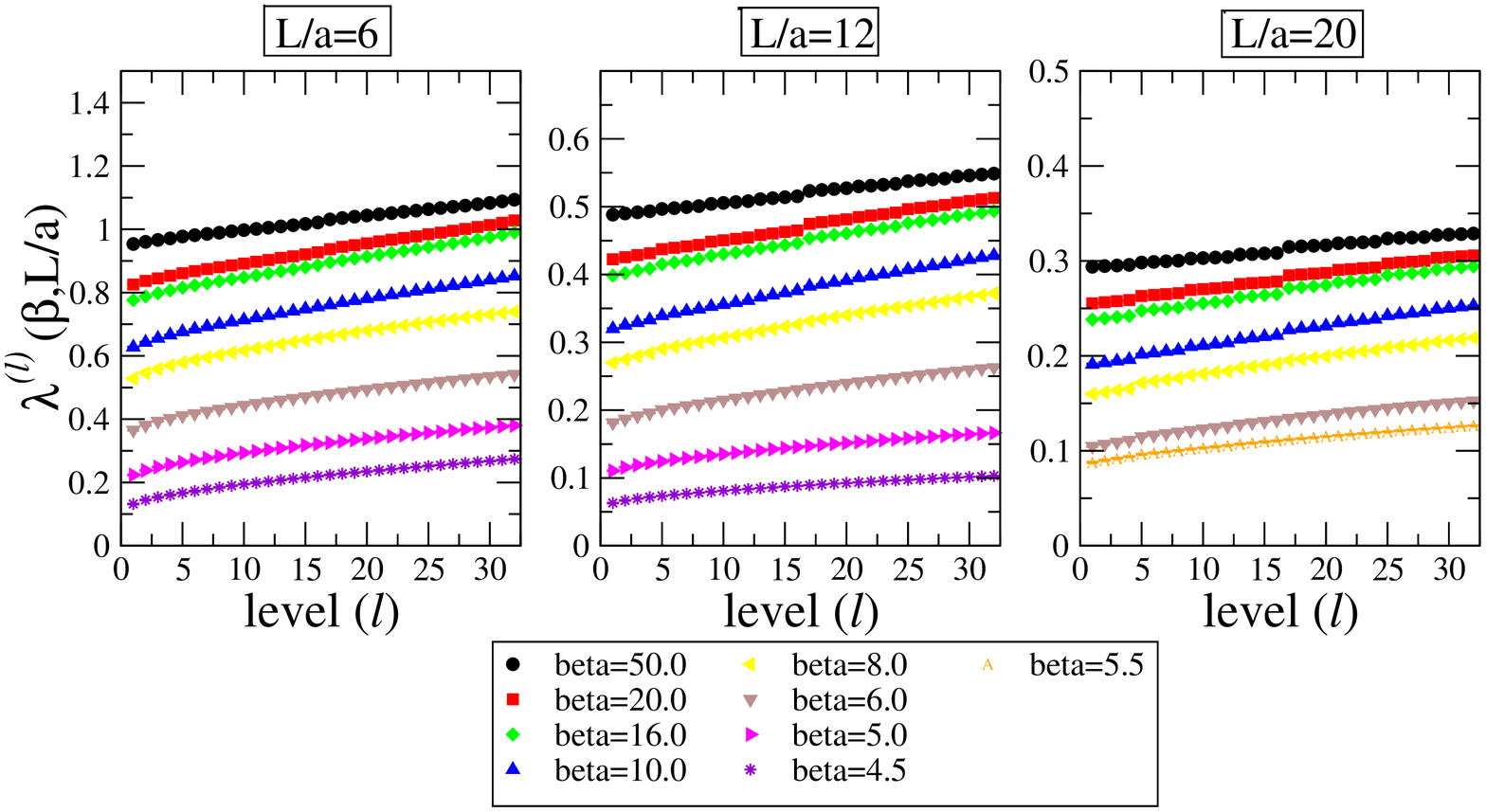} 
\caption{The examples of the eigenvalues for several $\beta$ and $L/a$. The statistical error is the same size of the symbol.}
\label{fig:eigenvalue-global}
\end{center}
\end{figure}

We also measure the eigenvalue for all lattice parameters in the data sets in Appendix~\ref{sec:app-Japan-data}.
We show some examples in Fig.~\ref{fig:eigenvalue-global}, in which the error is estimated by the bootstrap analysis.
The eigenvalues of high $\beta$ and the large lattice size show the $4$-fold degeneracy, although there is large taste breaking in the small lattice size or low $\beta$ region.
The data at the lowest $\beta$, $\beta=4.5$ for $L/a=6,12$ and $\beta=5.5$ for $L/a=20$, show the inconsistency with zero, and the $\beta$ dependence of the data at fixed lattice extent is smooth in whole $\beta$ region.
If we assume that the Banks-Casher relation, and the chiral symmetry is also preserving even in the lowest $\beta$ in our simulation parameter.

To see the taste breaking in our step scaling analysis in the strong coupling regime, we would like to show the eigenvalues in the continuum limit by taking the TPL renormalized coupling as a reference of the same physics:
\beq
L \cdot \lambda^{(l)}_{\mathrm cont} \equiv \left. \lim_{a \rightarrow 0} \hat{L} \cdot \lambda^{(l)} (\beta, L/a) \right|_{g^2_{\mathrm{TPL}}(L_0/L)=const},
\eeq
where the $(L \cdot \lambda_{cont})$ is dimensionless quantity and the scale $L$ is defined by the value of the renormalized coupling constant.
\begin{figure}[h]
\begin{center}
\includegraphics*[height=9cm]{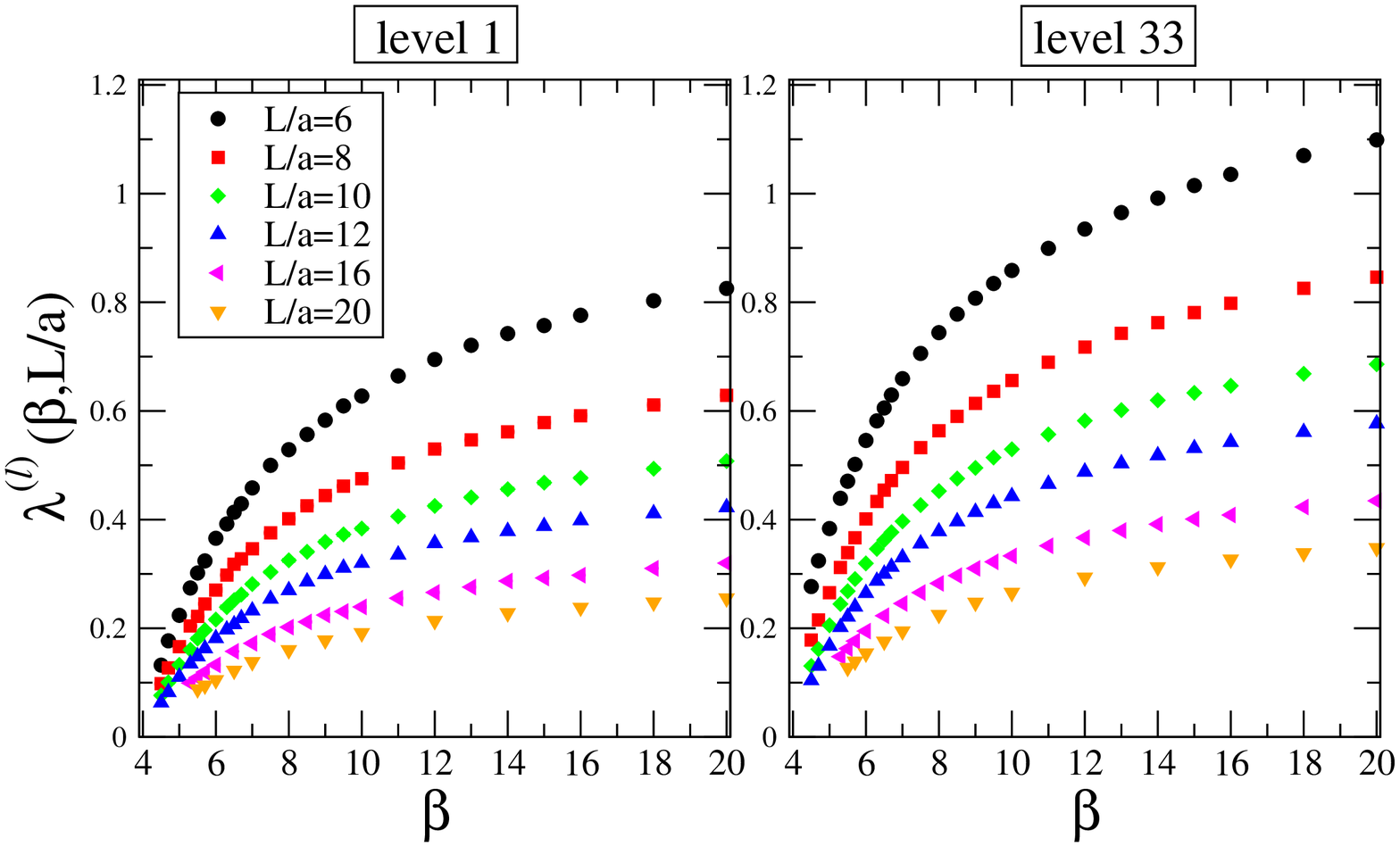} 
\caption{The $\beta$ dependence of the eigenvalues at the fixed $L/a$ for the level $1$ and $33$. The statistical error is the same size of the symbol.}
\label{fig:eigenvalue-beta-dep}
\end{center}
\end{figure}
The figures~\ref{fig:eigenvalue-beta-dep} show the $\beta$ dependence of the eigenvalues for the level $1$ and level $33$ for each lattice size.
We fit the data at fixed level and lattice size in terms of $\beta$ by the fitting function,
\beq
\hat{L} \cdot \lambda^{(l)} (\beta,L/a) =\sum_{i=0}^{N-1} c_i/ \beta^{i},
\eeq
where $N$ is the number of the fitting parameter, and in practical we choose the best fit value of $N$ for each $l$ and $L/a$.
Typically, $N=4$ -- $7$ are employed in this analysis.

In this analysis, the leading discretization error comes from the eigenvalue itself, which is proportional to $\hat{L} \cdot \lambda^{(l)} \propto  const. +O(a^2)$ if the theory lives in the deconfinement phase.
The other leading contribution $O(a^2)$ comes from the renormalized coupling, and the contribution comes via the tuned value of $\beta$.
We take the continuum limit for $5$ data points, $L/a=8,10,12,16$ and $20$, and they can be fitted well by the linear function of $(a/L)^2$ in whole region. 
To estimate the systematic uncertainty of this continuum extrapolation we also show the quadratic extrapolation of $(a/L)^2$ for $6$ data points included the coarsest lattice $L/a=6$.

\begin{figure}[h]
\begin{center}
\includegraphics*[height=8cm]{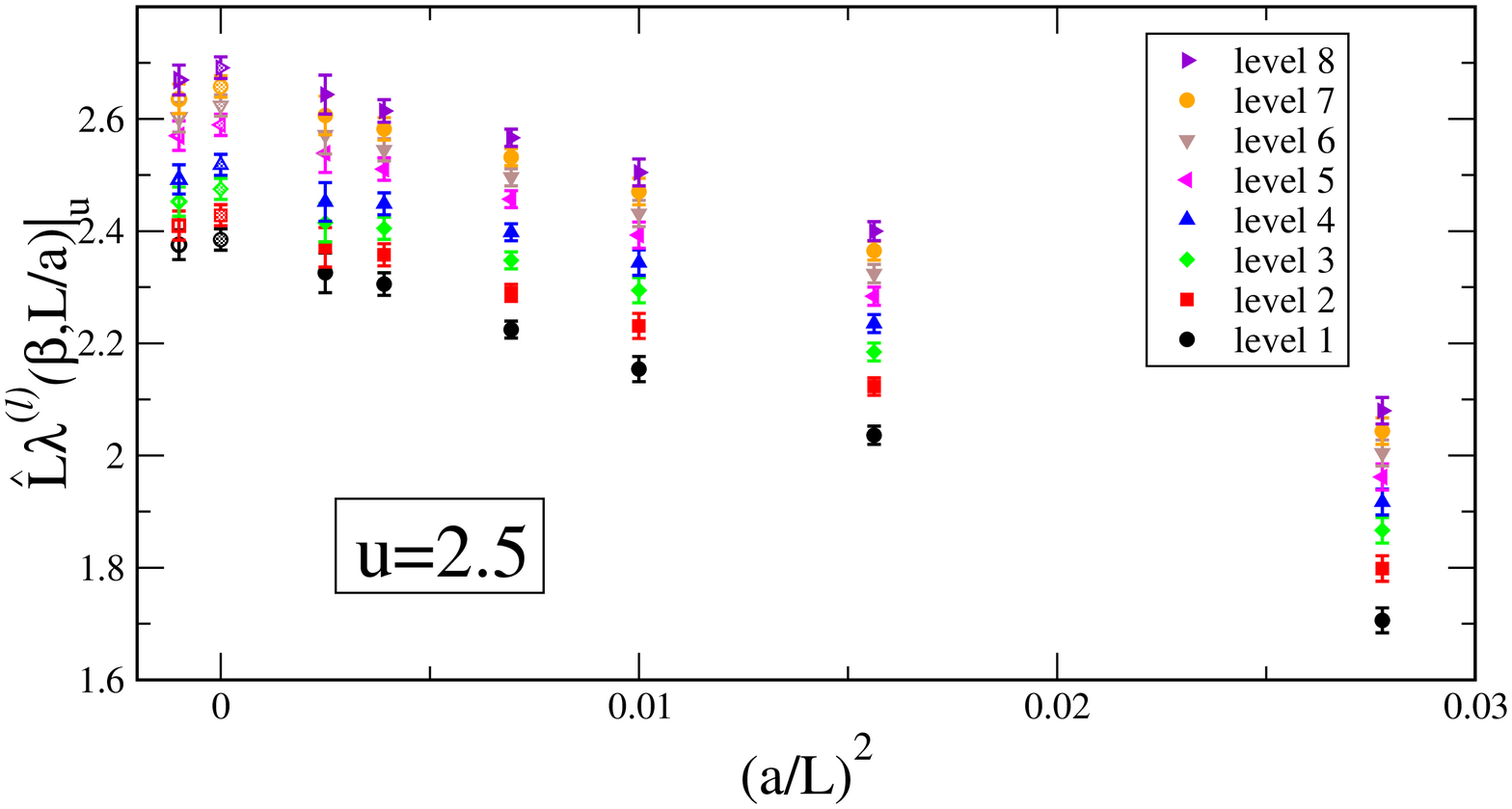} 
\caption{The continuum extrapolation for the low lying eigenvalues for $u=2.5$. Each full symbol denotes the eigenvalue at the interpolated $\beta$ for each lattice size; $L/a=6,8,10,12,16$ and $20$ from the right to left.
The shadow symbol at $(a/L)^2=0$ denotes the extrapolated value by the linear extrapolation function of $(a/L)^2$ for the finer $5$ lattice data. The empty symbol at $(a/L)^2=-0.001$ shows the extrapolated value for the quadratic function of $(a/L)^2$ by using all $6$ data points.}
\label{fig:eigenvalue-cont-lim}
\end{center}
\end{figure}
As an example, we take $u=2.5$, which corresponds to the region for $(\beta,L/a)=(5.4, 6)$ -- $(6.4, 20)$. 
The taste breaking of the raw data in these region is strong.
The continuum extrapolation for $u=2.5$ is shown in Fig.~\ref{fig:eigenvalue-cont-lim}.
In this plot, we show the eigenvalues at continuum limit for the low lying eigenvalues; level $1$ -- $8$.
The shadow symbol at $(a/L)^2=0$ denotes the extrapolated value for the linear extrapolation function of $(a/L)^2$ for the finer $5$ lattice data. The empty symbol at $(a/L)^2=-0.001$ shows the extrapolated value for the quadratic function of $(a/L)^2$ by using all $6$ data points.
The difference between these two kinds of the extrapolation can be identified as the systematic uncertainty.
Including the systematic error, we find that the breaking of the level $1$ -- $4$ becomes mild at the continuum limit even in the strong coupling regime.  
We also consider the order $a^4$ effect in the running coupling study (Fig.~\ref{fig:cont-lim-local}).
The recoveries of the taste breaking in the continuum limit using the extrapolation function at the same order would show an indirect evidence that we have considered sufficient order in $a^2$ in taking the continuum extrapolation of the TPL coupling constant.


\end{document}